\documentclass[aps,prl,epsfig,twocolumn,superscriptaddress]{revtex4-2}

\usepackage{graphics,graphicx, array, afterpage,breakurl,color,setspace}
\usepackage{qcircuit,amsmath}
\usepackage{grffile}
\usepackage[export]{adjustbox}
\usepackage{lineno}
\usepackage{xcolor}
\usepackage{longtable}
\usepackage{braket}
\usepackage{multirow}
\usepackage{enumerate}
\usepackage[colorlinks, citecolor=red]{hyperref}
\usepackage{amsmath,amssymb,amsfonts,amstext}
\usepackage{amsthm,times}
\usepackage[T1]{fontenc}
\usepackage[normalem]{ulem}
\usepackage{upgreek}
\usepackage{bm,framed}
\usepackage{mathrsfs}
\usepackage{multirow}
\usepackage{tabularx}

\begin{document}

\title{Experimental unsupervised learning of non-Hermitian knotted phases with solid-state spins}

\author{Yefei Yu}\thanks{These authors contributed equally to this work.}
\author{Li-Wei Yu}\thanks{These authors contributed equally to this work.}
\author{Wengang Zhang}\thanks{These authors contributed equally to this work.}
\author{Huili Zhang}
\author{Xiaolong Ouyang}
\author{Yanqing Liu}
\affiliation{Center for Quantum Information, IIIS, Tsinghua University, Beijing 100084, P. R. China}
\author{Dong-Ling Deng}
\email{dldeng@tsinghua.edu.cn}
\affiliation{Center for Quantum Information, IIIS, Tsinghua University, Beijing 100084, P. R. China}
\affiliation{Shanghai Qi Zhi Institute, 41th Floor, AI Tower, No. 701 Yunjin Road, Xuhui District, Shanghai 200232, China}

\author{L.-M. Duan}\email{lmduan@tsinghua.edu.cn}
\affiliation{Center for Quantum Information, IIIS, Tsinghua University, Beijing 100084, P. R. China}

\date{\today}


\begin{abstract}

Non-Hermiticity has widespread applications in quantum physics. It  brings about distinct topological phases without Hermitian counterparts, and gives rise to the fundamental challenge of phase classification from both theoretical and experimental aspects.  Here we report the  first experimental demonstration of unsupervised learning of  non-Hermitian  topological  phases  with  the nitrogen-vacancy  center platform. In particular, we implement the non-Hermitian twister model, which hosts  peculiar knotted topological phases,  with a solid-state quantum simulator consisting of an electron spin and a nearby $^{13}$C nuclear spin in a nitrogen-vacancy center in  diamond. By tuning the microwave pulses, we efficiently generate a set of  experimental data without phase labels. Furthermore, based on the diffusion map method, we  cluster this set of experimental raw data into three different knotted phases in an unsupervised fashion without {\em a priori} knowledge of the system,  which is in sharp contrast to the previously implemented supervised learning phases of matter. 
Our results showcase the intriguing potential for autonomous classification of exotic unknown topological phases with experimental raw data.

\end{abstract}
\maketitle

\begin{figure}
\centering
\includegraphics[width=\linewidth]{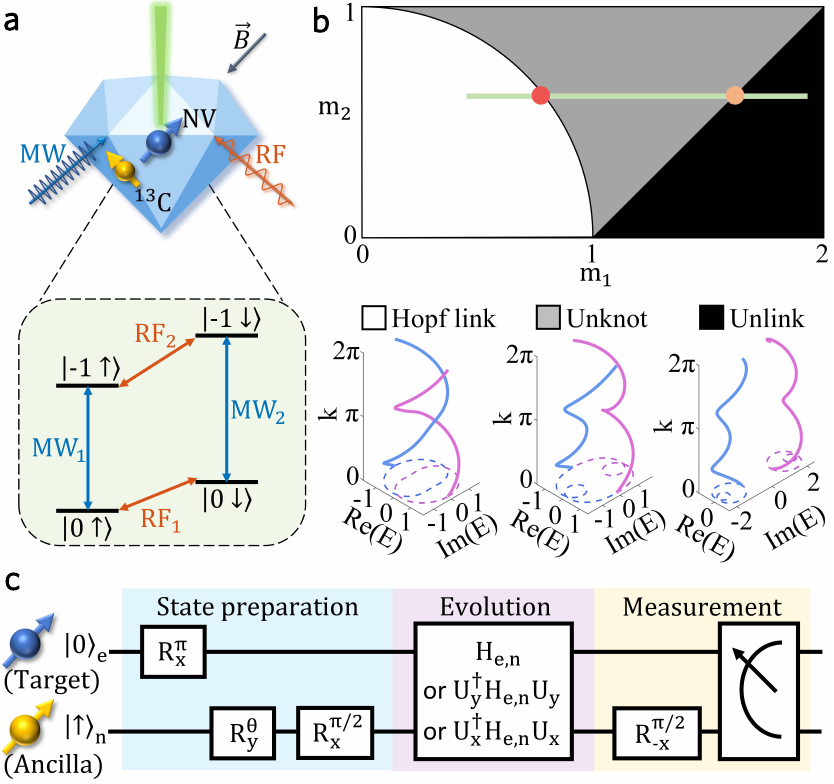}
\caption{
{ Simulating the non-Hermitian twister Hamiltonian with knotted topological phases on the NV center platform.}
{\bf a}, Top: schematic illustration of the NV center in diamond. The system is coherently controlled by the microwave (MW) and radio frequency (RF) pulses. The NV electron spin (blue arrow) is coupled to a nearby $^{13}$C nuclear spin (yellow arrow). Below: energy level diagram of the electron-nuclear spin system. MW controls the transitions between electron spin states \{$|0\rangle$, $|-1\rangle$\}, and RF controls the transitions between nuclear spin states \{$|\uparrow\rangle, |\downarrow\rangle$\}.
{\bf b}, Top: phase diagram of the non-Hermitian twister model.  We experimentally simulate the twister Hamiltonian with the parameters $m_{1,2}$ along the green line (crossing the two theoretical phase boundaries).  
Below: band structures of the twister model  in the space  spanned by $(\textrm{Re}(E), \textrm{Im}(E), k)$, the momentum $k$ is in the first Brillouin zone.  Gluing the  $k=0$ and $k=2\pi$ planes leads to different knotted band structures,  including the Hopf link (white square), the unknot (grey square), and the unlink (dark square).
{\bf c}, Quantum circuit for simulating the non-Hermitian Hamiltonian in our experiment. Here we take the electron (nuclear) spin as the target (ancillary) qubit. Through optical pumping,  we first polarize the state to be $|0\rangle_e|\uparrow\rangle_n$. By rotating along the x- and y- axes, we prepare the initial state $|\Psi(0)\rangle=|-1\rangle_e|-\rangle_n+\eta(0)|-1\rangle_e|+\rangle_n$. We then implement the unitary dynamics of the dilated Hamiltonian $H_{e,n}$, and measure the nuclear spin in the $|\pm\rangle_n$ bases. To measure the expectation value of $\sigma_{x,y}$, we apply a unitary transform to the target Hamiltonian: $\widetilde{H_e}=U_{y,x}^\dagger H_{e,n} U_{y,x}$. By postselecting the nuclear spin in the $|-\rangle_n$ state, one obtains the desired eigenstate of $H_e=H(k)$ for a given momentum $k$ for the electron spin (see Supplementary Information for details).
}
\label{mainfig1}
\end{figure}

Non-Hermiticity naturally emerges in a broad range of scenarios \cite{Moiseyev2011Non,Konotop2016Nonlinear,Ashida2020Nonhermitian}, 
and has been extensively studied in open quantum systems \cite{Rotter2009nonHermitian,Zhen2015Spawning,Diehl2011Topology,Verstraete2009Quantum}, photonics systems with loss and gain \cite{Feng2017NonHermitian,El-Ganainy2018Nonhermitian,Miri2019Exceptional, ozdemir2019parity,Ozawa2019Topological},  and quasiparticles with finite lifetimes \cite{Kozii2017NonHermitian,Zyuzin2018Flat,Shen2018Quantum,Zhou2018Observation,Yoshida2018NonHermitian}, etc.  
Recently, the interplay between the non-Hermiticity and topological phases has attracted considerable attentions \cite{Xu2017Weyl,Kunst2018Biorthogonal,Chen2018Hall,Lee2019Topological,Lee2016Anomalous, Luitz2015Manybody,Yoshida2018NonHermitian,Carvalho2018Realspace,Lee2019Anatomy,Leykam2017Edge,Yin2018Geometrical,Kawabata2019Topological,Gong2018Topological,Shen2018Topological,Yokomizo2019NonBloch,Ge2019Topological,Molina2018Surface,Xue2020NonHermitian,Budich2019Symmetryprotected,Yoshida2019Symmetry,Yang2019NonHermitiana,Okuma2020Topological,Li2020Critical,Yao2018Edge,Yao2018NonHermitian,Song2019NonHermitian,Yang2020Jones,Bessho2020Topological,Hockendorf2020Topological,Liu2019Secondorder,Deng2019NonBloch,Zeuner2015Observation,Poli2015Selective,Weimann2017Topologically,Chen2017Exceptional,Teo2010Majorana,Cerjan2019Experimental,Bandres2018Topological,Li2020Topological,Zhou2018Observation,Wang2021Topological,Hu2021Knots,Li2019Homotopical,Wojcik2020Homotopy,Kawabata2019Symmetry,Xiao2020NonHermitian,Weidemann2020Topological},  giving rise to an emergent research frontier of non-Hermitian topological phases of matter in both theory and experiment. Non-Hermitian topological phases bear a number of unique features without Hermitian analogs, 
including the non-Hermitian skin effect \cite{Yao2018Edge,Okuma2020Topological},  unconventional bulk-boundary correspondence \cite{Yao2018Edge}, and funneling of light \cite{Weidemann2020Topological}. 
To establish the theory of non-Hermitian topological phase classification, previous works have adopted the typical  homotopy-based approach akin to the Hermitian tenfold way,  and classified the non-Hermitian topological phases into 38 classes \cite{Kawabata2019Symmetry}. It was later recognized that the non-Hermitian topological phases can be further classified based on the knot or link structures of the complex energy bands, which gave rise to the new kind of knotted topological phases \cite{Wojcik2020Homotopy,Li2019Homotopical,Hu2021Knots}. 
 More recently, the braiding of such complex band structure has been implemented in experiment \cite{Wang2021Topological}. Yet,  hitherto it remains an ongoing challenge to completely classify the non-Hermitian topological phases from both theoretical and experimental aspects \cite{Kawabata2019Symmetry,Li2019Homotopical,Wojcik2020Homotopy,Helbig2020Generalized,Xiao2020NonHermitian,Weidemann2020Topological,Li2021Homotopical,Zhang2021Observation}. 



Machine learning methods provide an alternative and promising approach to classify  phases of matter \cite{Carrasquilla2017Machine}.
 Within the vein of learning topological phases, considerable strides have been  made from both theoretical  \cite{Zhang2017Quantum,Zhang2017Machine,Yoshioka2018Learning,Zhang2018Machine,Holanda2020Machine,Rodriguez-Nieva2019Identifying,Kottmann2020Unsupervised,Scheurer2020Unsupervised,Che2020Topological,Lidiak2020Unsupervised,Schafer2019Vector,Balabanov2020Unsupervised,Alexandrou2020Critical,Greplova2020Unsupervised,Arnold2021Interpretable} and experimental \cite{Lian2019Machine,Zhang2019Machine,Rem2019Identifying,Bohrdt2019Classifying,Kaming2021Unsupervised,Lustig2020Identifying} aspects, despite the fact that learning topological phases are more intricate than learning the symmetry-breaking ones due to  the lack of local order parameters   \cite{Beach2018Machine}. 
However, the above learning methods may not be straightforwardly extended to  the non-Hermitian scenario owing to the skin effect \cite{Yu2021Unsupervised}. This makes the machine learning non-Hermitian topological phases an intriguing task and a number of theoretical works, including both supervised and unsupervised methods,  have been proposed recently \cite{Yu2021Unsupervised,Long2020Unsupervised,Narayan2021Machine,Zhang2021Machine}. The supervised learning methods require  prior labelled samples, hence ruling out the capability of learning unknown phases. While the unsupervised learning can classify different topological phases from unlabelled raw data, without any prior knowledge about the underlying topological mechanism. Consequently, the unsupervised learning methods are more powerful in detecting unknown topological phases. One appealing unsupervised approach is based on the diffusion map \cite{Coifman2005Geometrica}, which has been theoretically demonstrated effective in clustering both Hermitian  \cite{Rodriguez-Nieva2019Identifying} and non-Hermitian  \cite{Yu2021Unsupervised} topological phases.
 However, to date the capability of machine learning methods in classifying non-Hermitian topological phases has not been demonstrated in experiment.

In this paper, we report the first experimental demonstration of  unsupervised learning of non-Hermitian knotted phases with a nitrogen-vacancy (NV) center in diamond. Specifically, we  utilize the dilation method \cite{Wu2019Observation,Zhang2021Observation} to  implement the desired non-Hermitian twister Hamiltonian with the NV center platform,   where  the electron spin constitutes the target system and a nearby $^{13}$C nuclear spin serves as  an ancilla  (Fig.~\ref{mainfig1}a).  Based on the non-unitary dynamics of the Hamiltonian with different parameters, we prepare an unlabelled data set (including 37 samples) with high fidelity by carrying out 3,552 non-unitary evolutions.  
Then we exploit the diffusion map method to cluster these experimental samples into different knotted topological phases in an unsupervised manner. 
The learning result matches precisely with the theoretical predictions, which clearly showcases the the robustness of the diffusion map method against the experimental imperfections. Besides, with the implemented set of samples, we experimentally realize different knot structures of the twister model, which can serve as the indices for different non-Hermitian topological phases.

 We consider the one dimensional (1D) non-Hermitian twister model under the periodic boundary condition, with the Hamiltonian taking the form \cite{Hu2021Knots},
\begin{equation}\label{Twister_Ham}
H(k)=\vec{d}(k)\cdot\vec{\sigma}=im_1\sigma_z+m_2 T_1+T_2,
\end{equation} 
where $\vec{d}(k)=(d_x,d_y,d_z)$, $\vec{\sigma}=(\sigma_x,\sigma_y,\sigma_z)$ are the Pauli matrices,  $k$ denotes the 1D momentum in the first Brillouin zone, $m_1$ and $m_2$ are tunable parameters (we set $\hbar=1$ for simplicity), 
and $T_n=\left[\begin{matrix}
0 & e^{ink}\\
1 & 0
\end{matrix}\right]$. 
This model hosts three distinct topological phases with phase boundaries $m_1^2+m_2^2=1$ and $m_2=\pm m_1-1$. In contrast to the typical homotopy-based approach for phase classification, these phases can  be efficiently classified by the knot (link) structures of the complex-energy bands (braid homotopy), where the knot structure is embedded in the space spanned by $(\textrm{Re}(E), \textrm{Im}(E), k)$, with $E$ denoting the complex energy. Concretely, the three non-Hermitian topological phases of the twister model are indexed by the Hopf link, the unlink, and the unknot,  respectively. The phase transition occurs when the knot structures of the complex bands change across the exceptional points \cite{Hu2021Knots}. It is worth noting that all of the three  phases host the non-Hermitian skin effect since the corresponding bands have the point gaps, which indicate that the phase transition points are boundary condition sensitive. A sketch of the phase diagram of the non-Hermitian twister model is shown in Fig.~\ref{mainfig1}b.

Experimentally simulating the non-Hermitian Hamiltonian is challenging,  since the dynamical evolution of closed systems is usually governed by the Hermitian Hamiltonians.  
One fruitful approach is to dilate the non-Hermitian Hamiltonian into Hermitian ones in a larger Hilbert space. The dilation method was  theoretically proposed to simulate the $\mathcal{PT}$--symmetric non-Hermitian Hamiltonian \cite{Kawabata2017Information}. 
Then it was applied in experiments to study the $\mathcal{PT}$--symmetry breaking and implement the non-unitary dynamics of photons \cite{Xiao2019Observation,Wu2019Observation}. More recently, this method is exploited for  simulating the dynamics of non-Hermitian Su-Schrieffer-Heeger band model with topological phases \cite{Zhang2021Observation}. 

\begin{figure*}
\centering
\includegraphics[width=1.8\columnwidth]{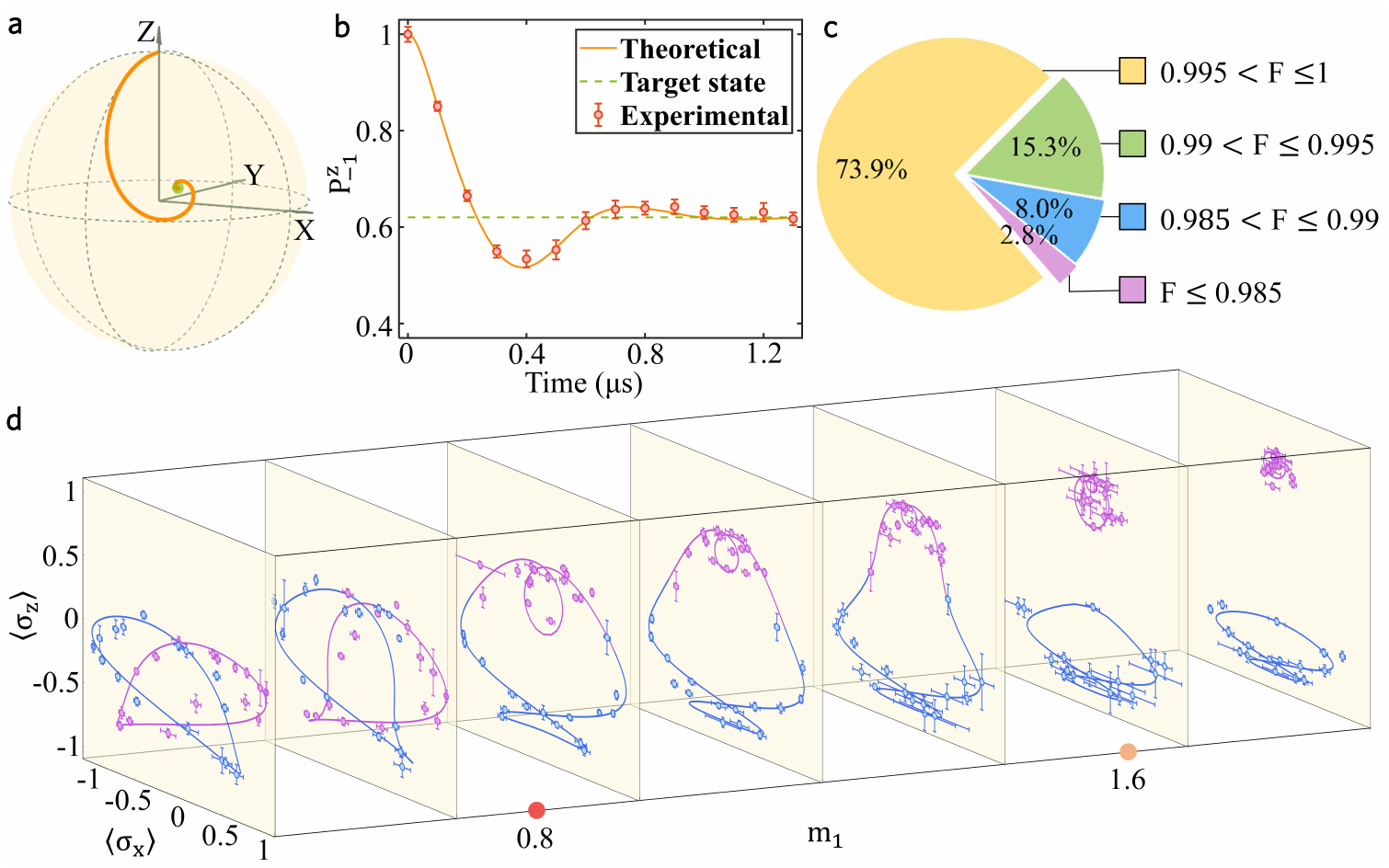}
\caption{
{ State preparation through non-unitary dynamics and observation of knot structures for different phases.} 
{\bf a},  Decaying  trajectory of the electron spin state on the Bloch sphere under non-unitary dynamics. The green dot represents the target state. 
{\bf b}, Non-unitary time evolution of the electron spin population $P_{-1}^{z}={\rm Tr}(\rho_{e}|-1\rangle_{e}\langle-1|)$. The red dots with error bars are the experimental results. Error bars are obtained via Monte Carlo simulation by assuming a Poisson distribution of the photon counts for 10,000 times. The orange solid line shows the theoretical curve predicted by numerical simulation. The green dashed line indicates the ideal population of the target state assuming infinite evolution time. For {\bf a} and {\bf b}, the underlying Hamiltonian is $-H(k)$ with the parameters $m_{1}=0.9855, m_{2}=0.6$, and $k=0.125\pi$.
{\bf c},  Pie chart illustrates the percentage distribution for the fidelity (denoted as F) of the 1,184 experimentally prepared eigenstates  $\{|R_1\rangle,|R_2\rangle\}$
with varying $m_{1}$ and $k$. 
{\bf d},  Experimental results for  $\langle\sigma_z\rangle$ versus $\langle\sigma_x\rangle$ as $k$ sweeping the first Brillouin zone with different values of $m_1$, where the blue and purple parts denote the average values of $\sigma_{x,z}$  with respect to $|R_1\rangle$ and $|R_2\rangle$, respectively. The dots with error bars represent the experimental data, whereas the solid lines denote the  theoretical predictions. The red and orange dots denote the phase transition values of $m_1$,  while fixing $m_2=0.6$. 
}
\label{mainfig2}
\end{figure*}

Here, we utilize the dilation method (see Methods and Supplementary Information) to implement the twister Hamiltonian $H(k)$. 
With such a dilation method, the simulation of the non-Hermitian  $H_e=H(k)$ for the electron spin is mapped to the simulation of a Hermitian Hamiltonian $H_{e,n}$ for the coupled electron and nuclear spins. Fig.~\ref{mainfig1}c illustrates the  quantum circuit  for our  experiment. To implement $H_{e,n}$, we apply two microwave pulses with time-dependent frequency, amplitude and phase. Then based on the non-unitary dynamics governed by $H_e$ in the electron spin subsystem, we  explore how the state evolves to the desired eigenstate of $H(k)$ by checking the electron spin population on $|-1\rangle_e$ (see Fig.~\ref{mainfig2}a for a schematic  demonstration). Fig.~\ref{mainfig2}b displays our experimental results of the electron spin population on the state $|-1\rangle_e$, and shows that the experimental results coincide with the theoretical predictions very well, with almost all of the data points being within the error bars. After long time evolution ($\sim$1.2$\mu s$), the electron spin state decays to the targeted eigenstate of $H(k)$. This shows that the dilation method is effective and efficient in simulating  the non-Hermitian  twister model.

Typically, the success of the diffusion map method relies crucially on the data samples. To learn non-Hermitian topological phases in an unsupervised fashion,  one candidate data set is the bulk Hamiltonian unit vectors \cite{Yu2021Unsupervised}:   ${\bf X}=\{{\bf x}^{(l)}| {\bf x}^{(l)} = \{\frac{1}{N}\hat{\bf d}(k_i),| k_i =\frac{2i-N-2}{N}\pi, \, i\in[1,N] \}\}$ with $\hat{\bf d} = \frac{ \vec{d}}{\sqrt{d_x^2+d_y^2+d_z^2}}$ and $N$ denotes the number of unit cells of the system. To fit the experimentally prepared data (the right eigenstates $|R_1\rangle$ and $|R_2\rangle$) into the diffusion map algorithm, we transform the states $|R_{1,2}\rangle$ into the unit Hamiltonian vector: 
\begin{equation}
\hat{d}_{x,y,z}=\frac{1}{2}\text{Tr}[\left\{(|R_1\rangle\langle L_1|-|R_2\rangle\langle L_2|), \sigma_{x,y,z}\right\}],
\end{equation}
where the curly brackets denote the anti-commutator, the left and right eigenstates of the Hamiltonian  obey the biorthogonal relation $\langle L_\alpha|R_\beta\rangle=\delta_{\alpha\beta}$ with $\alpha,\beta\in \{1,2\}$, and $\delta$ denoting the Kronecker delta function. Therefore, once we obtain the right eigenvectors $|R_1\rangle$ and $|R_2\rangle$  of $H(k_i)$ in the experiment, we can derive the other two eigenstates $|L_1\rangle$ and $|L_2\rangle$. Then by varying the discrete momentum $k_i$ in the first Brillouin zone with step-size $\pi/8$ while fixing $m_1$ and $m_2$, we can prepare one experimental data sample.  Consequently, we obtain the experimental data set  of 37 samples by varying  $m_1$, 
with fixed $m_2=0.6$. 
From Fig.~\ref{mainfig2}c, we see that more than $97.2\%$ of the 1,184 prepared states have a fidelity larger than 0.985, which indicates the high quality of our prepared data and the accurate controllability of the system. 

One can also probe the topological phase transition of the non-Hermitian twister model by measuring $\langle\sigma_x\rangle$ and $\langle\sigma_z\rangle$ with respective to $|R_1\rangle$ and $|R_2\rangle$, with $k$ sweeping the first Brillouin zone. Fig.~\ref{mainfig2}d displays the experimental results of the trajectories of $\langle\sigma_x\rangle$ and $\langle\sigma_z\rangle$ with  varying  $m_1$. 
The trajectories of $\langle\sigma_x\rangle$ and $\langle\sigma_z\rangle$ form three types of structures, namely the two overlapping closed loops ($m_1<0.8$), one closed loop ($0.8<m_1<1.6$), and two separate closed loops ($m_1>1.6$), corresponding respectively  to the Hopf link, unknot, and unlink topological phases. This coincides exactly with the theoretical prediction. Based on the experimental data, we can also calculate the global biorthogonal Berry phases that are related to the parity of band permutations \cite{Hu2021Knots} (see Table S1 of Supplementary Information).



In Fig.~\ref{DM_EXP_THE}, we show the unsupervised learning results for the twister Hamiltonian based on both the numerically simulated and   experimental data sets, respectively.
Fig.~\ref{DM_EXP_THE}c shows the kernel value distribution with experimental data samples, where the samples belonging to the same yellow block can diffuse to each other with a sizable probability, and hence can be clustered together based on the connectivity. As a consequence,  the experimental data samples are clustered into three categories in the dimension-reduced space, see Fig.~\ref{DM_EXP_THE}d. In addition, we can obtain the two phase boundaries from Fig.~\ref{DM_EXP_THE}c, which agree precisely with the theoretical prediction, as well as the boundaries of  phases indexed by the experimental trajectory results in Fig.~\ref{mainfig2}d.
We note that the 29th sample with $m_1=1.6015$ is very close to the phase transition point ($m_1=1.6$), which causes a large deviation from its presumed category (see the grey circles in Fig.~\ref{DM_EXP_THE}b and Fig.~\ref{DM_EXP_THE}d).
Besides, 
by comparing our experimental result with the numerical one in Fig.~\ref{DM_EXP_THE}a,b, we obtain that the diffusion map method is sufficiently robust against the experimental noises.

\begin{figure}
\centering
\includegraphics[width=\linewidth]{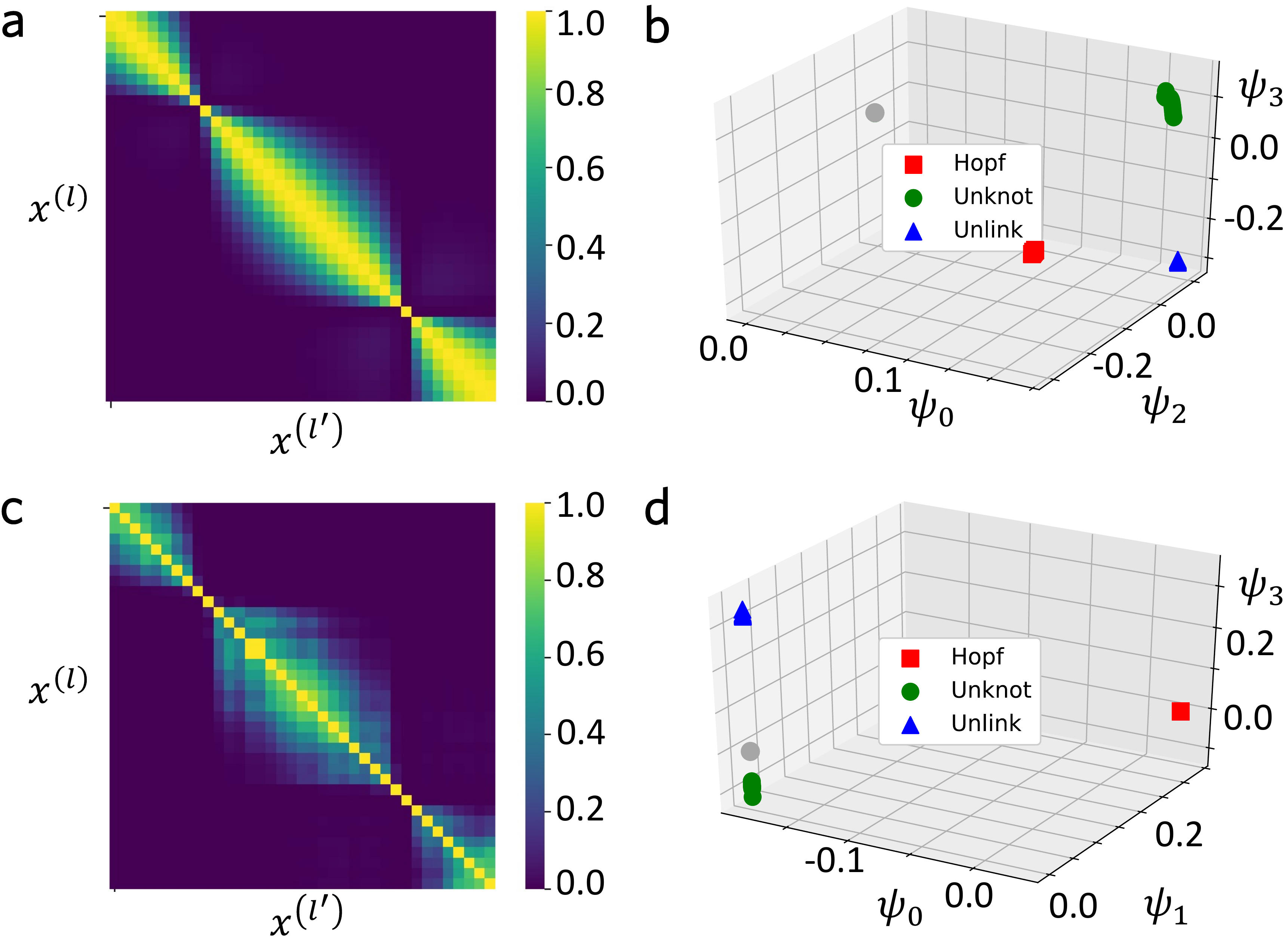}
\caption{
{Unsupervised clustering results for the knotted topological phases with the numerically simulated and experimental data sets. 
} 
{\bf a}, Heatmap for Gaussian kernel value distribution between numerically simulated samples with varying $m_1$. Samples belonging to the same yellow block can diffuse to each other with non-zero probability, and hence are clustered together based on the connectivity. The dividing points between the yellow blocks  correspond to the phase boundaries.
{\bf b}, Scatter diagram of the three eigenvectors of the diffusion matrix with the corresponding eigenvalues $\lambda\approx1$. The numerically simulated data samples are clustered into three knotted topological phases, where the red squares, green circles, and blue triangles denote the Hopf link, unknot, and unlink phases, respectively. 
{\bf c}, Heatmap for Gaussian kernel value distribution between experimental samples.  
{\bf d}, Scatter diagram of the dimension-reduced experimental data samples. The experimental data samples are clustered into three topological phases, which agree precisely with the numerical simulation. The grey circles in {\bf b} and {\bf d} denote the 29th sample with $m_1=1.6015$, which is very close to the theoretical phase transition point $m_1=1.6$. This leads to the pronounced deviation from its presumed category. 
Parameters are chosen as: the variance parameter $\epsilon=0.08$, $m_2=0.6$,  the varying parameter $m_1^{(l)}=0.4106+l*4/\pi^4$ for each sample ${\bf x}^{(l)}$, with $l\in [1,37]$. 
}
\label{DM_EXP_THE}
\end{figure}

We emphasize that the applicability of the diffusion map method in classifying non-Hermitian topological phases can be explained from a physical perspective. On the one hand, it has been rigorously proved that non-Hermitian samples divided by the band crossing points cannot be clustered together via diffusion maps \cite{Yu2021Unsupervised}. 
On the other hand, a more general theory for the non-Hermitian topological phase classification  only assumes separable bands \cite{Shen2018Topological}. For the twister model, we have experimentally observed that the band crossing leads to a change of the knot structure (see Fig.~\ref{mainfig2}d), which indicates the transition between different topological phases. Hence from the perspective of separable bands, the diffusion map method matches naturally with the mathematical protocol for classifying non-Hermitian topological phases in the momentum space. We also note that the non-Hermitian twister model bears unconventional bulk-boundary correspondence, which means that the phase boundaries are sensitive to the boundary conditions. In the future, it would be interesting and important to experimentally demonstrate the unsupervised learning of non-Hermitian topological phases under the open boundary condition. Achieving  this requires meticulous and accurate engineering of many spin interactions, which is still a notable challenge with the current NV technologies.

To summarize, we have demonstrated unsupervised learning of non-Hermitian topological phases, based on the non-unitary dynamical evolution of the electron spin with the NV center platform. In particular, we have generated a high-fidelity experimental data set of the non-Hermitian twister model and successfully clustered these experimental samples into different knotted phases in an unsupervised fashion.  
Our work paves a way to use unsupervised machine learning  to identify undiscovered non-Hermitian topological phases with the state-of-the-art experimental platforms.

We acknowledge helpful discussions with Z. Wang and S.  Zhang. This work was supported by the Frontier Science Center for Quantum Information of the Ministry of Education of China, Tsinghua University Initiative Scientific Research Program, and the Beijing Academy of Quantum Information Sciences, and the National Natural Science Foundation of China (Grants No. 12075128 and No. 11905108). D.-L.D. also acknowledges additional support from the Shanghai Qi Zhi Institute.

\bibliography{DengQAIGroup}

\clearpage
\onecolumngrid
\makeatletter
\setcounter{figure}{0}
\setcounter{equation}{0}

%



\setcounter{secnumdepth}{3}

\makeatletter
\renewcommand{\thefigure}{S\@arabic\c@figure}
\renewcommand \theequation{S\@arabic\c@equation}
\renewcommand \thetable{S\@arabic\c@table}


\makeatletter
\renewcommand{\thefigure}{S\@arabic\c@figure}
\renewcommand \theequation{S\@arabic\c@equation}
\renewcommand \thetable{S\@arabic\c@table}

\begin{center} 
	{\large \bf Supplementary Information: Experimental unsupervised learning of non-Hermitian knotted phases with solid-state spins}
\end{center}

\renewcommand\thesection{S\Roman{section}}
\tableofcontents

\newpage
\section{The diamond sample and experimental setup}
\begin{figure}[h]
	\centering
    \includegraphics[width=0.95\linewidth] {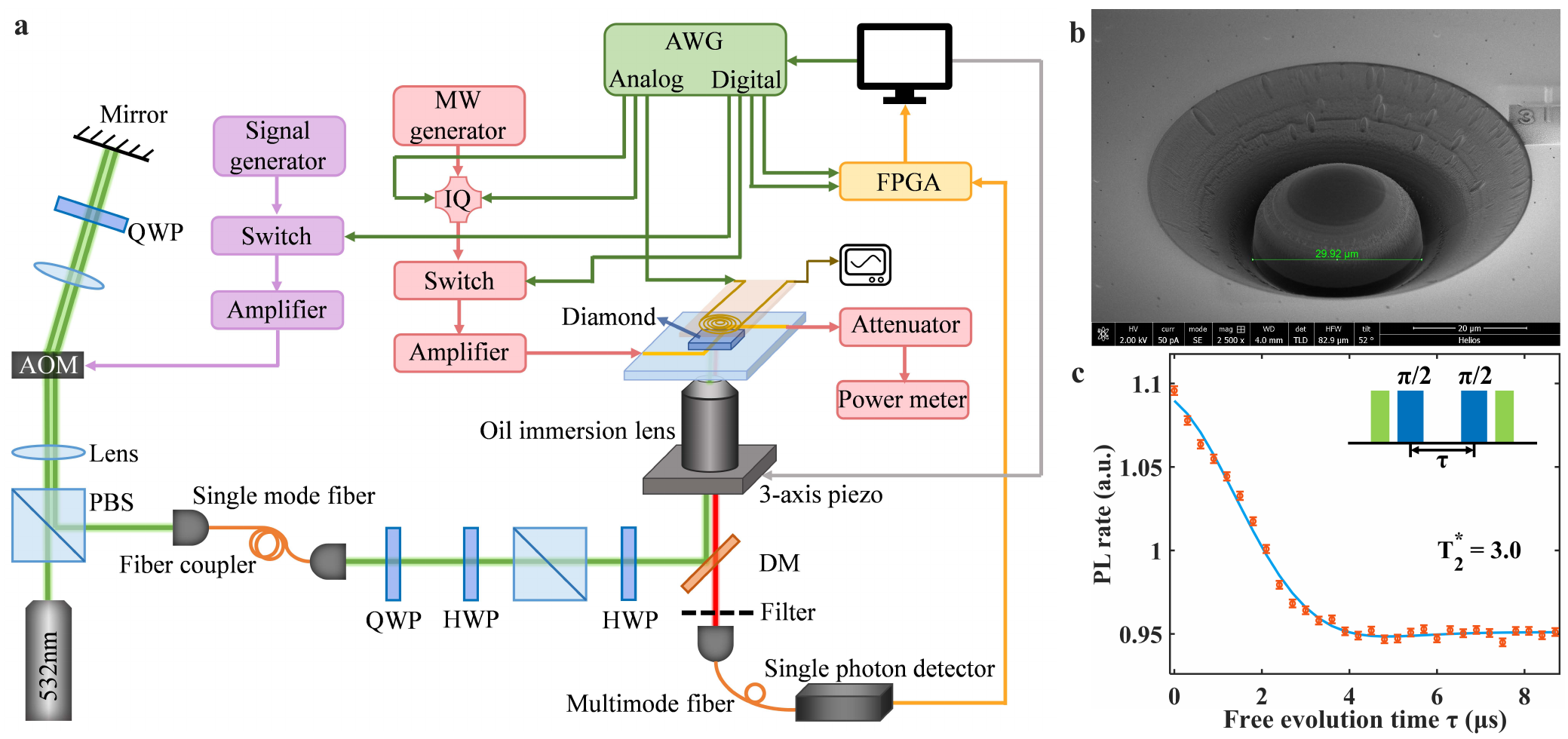}
    \caption{\textbf{ The diamond sample and experimental setup.}  
 {\bf a}, Schematic diagram of our experimental setup. 
 {\bf b}, Image of a solid immersion lens (SIL) under scanning electron microscope (SEM). The SIL is fabricated by the focused ion beam (FIB).
 {\bf c}, We measure the dephasing time of the electron spin via standard Ramsey interferometry. The experimental data (red circle) is fitted with $f(\tau)=a+b\cos(\delta t+\varphi)\exp(-(\tau/T_2^*)^2)$ (solid blue line), giving $T_2^* = 3.0$ $\mu s$. 
}
\label{Supp_Exp_Setup}
\end{figure}

Our experiments are performed on a $\langle 100 \rangle$-oriented single crystal diamond (type $\uppercase\expandafter{\romannumeral2}$a) produced by Element Six with a natural abundance of carbon isotopes ([$^{13}$C]=$1.1\%$). We utilize a single NV center with a neighboring $^{13}$C atom of $13.7$ MHz hyperfine strength. 
A solid immersion lens (SIL) is fabricated on top of the preselected NV center to enhance the collection efficiency (see Fig. \ref{Supp_Exp_Setup}b). The photoluminescence rate of the NV center is about $460$ kcps under $80$ $\mu W$ laser excitation.\\

The diamond sample is mounted on a confocal microscopy system (see Fig. \ref{Supp_Exp_Setup}a). A $532$ nm green laser is used for spin state initialization and readout. The laser beam is then modulated by an acoustic optical modulator (AOM, ISOMET 1250C-848) to generate laser pulses. To avoid continuous polarization caused by the laser leakage, the first-order diffracted beam generated by AOM is reflected by a mirror, forming a double-pass structure to enhance the on-off ratio to $10^5$:$1$. The green laser is coupled to a single-mode fiber, guided out by a collimator,  reflected by a dichroic mirror (DM), and then  focused on  our sample  through an oil-immersion objective lens. 
The fluorescence photons of the NV center are collected via the same objective lens and pass through the DM followed by a $637$ nm long-pass filter. Then the photons are coupled to a multi-mode fiber and detected by a single photon detector module (SPDM). A homemade field-programmable gate array (FPGA) board is applied to count the fluorescence photons.\\ 

In order to coherently manipulate the electron spin and nearby nuclear spins,
we use an arbitrary waveform generator (AWG, Techtronix 5014C) to generate transistor-transistor logic (TTL) signals and low frequency analog signals. One of the TTL signal controls the on/off of AOM, namely the laser pulses. Another two TTL signals provide gate signals for the FPGA board. For the manipulation of the electron spin, the carrier MW signal generated by a MW source (Keysight N5181B) is combined with two of the analog signals of AWG through an IQ-mixer (Marki Microwave IQ1545LMP). For the manipulation of the nuclear spin, another analog signal is used to generate the RF signal. Both MW and RF signals are further amplified by amplifiers (Mini Circuits ZHL-30W-252-S+ for MW and Mini Circuits LZY-22+ for RF). The MW signal is then delivered to the diamond sample via a gold coplanar waveguide (CPW). The RF signal is applied through a homemade copper coil.\\ 

All the experiments in this work are implemented at the room temperature. To polarize the nuclear spins via excited-state level anticrossing (ESLAC) \cite{Smeltzer2009Robust}, a static magnetic field of $B_{z} \approx 480$ Gauss is applied along the NV axis by a permanent magnet. To ensure that the coherence time is sufficient for the time evolution process ($\sim 1\,\mu s$) in our experiments, we perform a Ramsey interferometry measurement (see  Fig. \ref{Supp_Exp_Setup}c). For the NV center used in this work, the coherence time $T_2^*$ is measured to be $3.0$ $\mu s$. 

\newpage
\section{Implementation of the non-Hermitian Hamiltonian with the NV center platform}
\subsection{Details of the dilation method for realizing non-Hermitian Hamiltonian}
In this work, we focus on experimentally simulating the  non-Hermitian twister Hamiltonian $H(k)$ \cite{Hu2021Knots},
\begin{equation}\label{Supp_Twister_Ham}
H(k)=\vec{d}\cdot\vec{\sigma}=im_1\sigma_z+m_2 T_1+T_2,
\end{equation} 
where $\vec{d}=(d_x,d_y,d_z)$, $\vec{\sigma}=(\sigma_x,\sigma_y,\sigma_z)$ are the Pauli matrices,  $k$ denotes the 1D momentum in the first Brillouin zone, $(m_1,m_2)$ are tunable parameters (we set $\hbar=1$ for simplicity), 
and $T_n=\left[\begin{matrix}
0 & e^{ink}\\
1 & 0
\end{matrix}\right]$. 
It is generally difficult to straightforwardly simulate the non-Hermitian Hamiltonian with a closed quantum system. To tackle such an issue,  we adopt the dilation method to extend the target non-Hermitian system into a larger Hilbert space governed by a Hermitian Hamiltonian.   

 Here,  we present the corresponding mathematical details. The dynamical evolution of the target electron spin state  $|\psi(t)\rangle_e$ in the non-Hermitian system $H_e$ is described by  the Schr{\"o}dinger equation $ i \frac{\partial }{\partial t} |\psi(t)\rangle_e= H_e |\psi(t)\rangle_e$.

To perform the dilation method, one need to introduce an ancilla qubit (nuclear spin) into the system, with the   dilated state denoted by $|\Psi(t)\rangle_{e,n} = |\psi(t)\rangle_e|-\rangle_n+\eta(t)|\psi(t)\rangle_e |+\rangle_n$, where $|-\rangle_n = (|\uparrow\rangle_n-i|\downarrow\rangle)_n/\sqrt{2}$ and $|+\rangle_n = -i(|\uparrow\rangle_n+i|\downarrow\rangle_n)/\sqrt{2}$ are the eigenstates of Pauli matrix $\sigma_y$ in the ancilla space, $\eta(t)$ is a time-dependent operator to be determined. 
 In such dilated system, the state $|\Psi(t)\rangle_{e,n}$  evolves under a Hermitian Hamiltonian $H_{e,n}$, with the corresponding Schr{\"o}dinger equation  described by $i \frac{\partial }{\partial t} |\Psi(t)\rangle_{e,n}= H_{e,n} |\Psi(t)\rangle_{e,n}$. 

In Ref. \cite{Wu2019Observation}, it has been proved that the dilation method can be utilized for realizing all of the Hermitian and non-Hermitian Hamiltonians,  which means that one can always find a dilated Hermitian Hamiltonian $H_{e,n}$ for any target non-Hermitian $H_e$. The dilated Hamiltonian $H_{e,n}$ takes the following general formula \cite{Wu2019Observation},
\begin{equation}
H_{e,n}=\mathcal{Q}(t)\otimes\sigma_z+\mathcal{P}(t)\otimes \mathbf{I},
\label{DH}
\end{equation}
with the Hermitian operators 
$\mathcal{Q}(t)= i[H_e\eta(t)-\eta(t)H_e-i\frac{d}{dt} \eta(t)]\mathcal{R}^{-1}(t)$,
$\mathcal{P}(t)=\{H_e+[i\frac{d}{dt} \eta(t)+\eta(t)H_e]\eta(t)\}\mathcal{R}^{-1}(t)$,
$\mathcal{R}(t)=\eta^\dagger(t)\eta(t)+\mathbf{I}$. 
The time-dependent operator $\mathcal{R}(t)$ should satisfy the following dynamical equation
\begin{equation}
i\frac{d}{dt}\mathcal{R}(t)=H_e^\dagger \mathcal{R}(t)-\mathcal{R}(t)H_e.
\end{equation}
We remark that the initial condition $\mathcal{R}(t=0)$ (namely $\eta(0)$) should be judiciously chosen, so as to keep $\mathcal{R}(t)-\mathbf{I}$ positive throughout our experiment.

\subsection{Realization of the dilated Hermitian Hamiltonian with the NV center system}
Here,  we present more details about how to experimentally realize the dilated Hamiltonian $H_{e,n}$ in our NV center system. We first give a brief introduction to the Hamiltonian of NV center system. The NV center is a point defect in the diamond lattice, where two adjacent carbon atoms are replaced by a nitrogen atom and a vacancy. A negatively charged $NV^{-}$ has electron spin triplet ($S=1$) ground states. Now we consider a system consisting of an NV center electron spin and a strongly coupled $^{13}$C nuclear spin. By applying an external magnetic field $B$ along the NV axis and using secular approximation, one can express the Hamiltonian of the two-qubit system  as
\begin{equation}
H_{\text{NV}}=   A_\text{zz}S_zI_z+\gamma_n B I_z +\gamma_e B S_z +DS_z^2, \label{NVHam}
\end{equation}
where $I_z$ denotes the $z$-component of the nuclear spin $S=1/2$, $S_z$ denotes the $z$-component of the total electron spin $S=1$,  
$\gamma_e$ ($\gamma_n$) represents the gyromagnetic ratio of the electron (nuclear) spin, $D=2\pi\times 2.87$ GHz denotes the zero-field splitting of electron spin, and $A_\text{zz}=2\pi\times 13.7$ MHz represents the hyperfine coupling strength between the electron and nuclear spins. Here, to realize the two-band target system  in our experiment,  we only choose two of the electron ground states, $|0\rangle_e$ and $|-1\rangle_e$, as shown in Fig. 1\text{b} of the main text. Thus,  the dilated Hilbert space is spanned by the two-qubit bases $\{|0\rangle_e|\uparrow\rangle_n, |0\rangle_e|\downarrow\rangle_n, |-1\rangle_e|\uparrow\rangle_n, |-1\rangle_e|\downarrow\rangle_n\}$.  In such two-qubit space,  the Hamiltonian in Eq. \ref{NVHam} reduces to the following formula:
\begin{equation}
H_{NV}\longrightarrow H_0=\tfrac{1}{2}A_\text{zz} \sigma_z\otimes \sigma_z-(D-\gamma_e B-\tfrac{1}{2}A_\text{zz})\sigma_z \otimes \mathbf{I}+(\gamma_n B-\tfrac{1}{2}A_\text{zz})\textbf{I} \otimes \sigma_z.
\label{NV_Hamiltonian}
\end{equation}

Then we apply two selective MW pulses to drive the two different electron spin transitions (Fig.1\text {b} in the main text). The corresponding Hamiltonian  $H_{\text{MW}}$  takes the form
\begin{equation}
\begin{split}
H_{\text{MW}}=\left(\sum_{i=1}^2 \pi\Omega_i(t)\cos [ \int_{0}^{t}\omega_i(\tau)d\tau +\phi_i(t)] \right)*\left(\sigma_x \otimes \mathbf{I}\right), 
\end{split}
\end{equation}
where $\Omega_i$, $\omega_i$ and $\phi_i$ denotes the amplitude, frequency and phase of the two microwaves $\text{MW}_{i}$ (Fig.1a of the main text), $i=1,2$. With the MV pluses, the total Hamiltonian $H_{\rm total}$ of the system becomes
\begin{equation}
H_{\rm total}= H_0+H_{MW}.
\end{equation}
With the total Hamiltonian $H_{\rm total}$, one can then realize the desired dilated Hamiltonian $H_{e,n}$ in the rotating frame.  


To enter the rotating frame, we insert a unitary matrix $U_{\text{rot}}$ on both sides of the Schr{\"o}dinger equation $i \frac{\partial |\varphi\rangle}{\partial t} = H_{\rm total} |\varphi\rangle$ to have
\begin{equation}
i  U_{\text{rot}} \frac{\partial |\varphi\rangle}{\partial t} = U_{\text{rot}} H_{\rm total} U_{\text{rot}}^{\dagger} U_{\text{rot}} |\varphi\rangle.
\end{equation}
Here we define $|\tilde{\varphi}\rangle = U_{\text{rot}}|\varphi\rangle$, from the property of differential we have
\begin{equation}
\frac{\partial |\tilde{\varphi}\rangle}{\partial t} = \frac{\partial (U_{\text{rot}}|\varphi\rangle)}{\partial t}=U_{\text{rot}}\frac{\partial |\varphi\rangle}{\partial t}+\frac{\partial U_{\text{rot}}}{\partial t}|\varphi\rangle.
\end{equation}
Thus, the state $|\tilde{\varphi}\rangle$ satisfies the following equation
\begin{equation}
    i\frac{\partial |\tilde{\varphi}\rangle}{\partial t}= U_{\text{rot}} H_{\rm total} U_{\text{rot}}^{\dagger}|\tilde{\varphi}\rangle-iU_{\text{rot}}\frac{\partial U_{\text{rot}}^{\dagger}}{\partial t} |\tilde{\varphi}\rangle =H_{\rm eff} |\tilde{\varphi}\rangle,
\end{equation}
where $H_{\rm eff}$  represents the effective Hamiltonian in the rotating frame, as
\begin{equation}
 H_{\rm eff} = U_{\text{rot}} H_{\rm total} U_{\text{rot}}^{\dagger}-iU_{\text{rot}}\frac{\partial U_{\text{rot}}^{\dagger}}{\partial t}.
\end{equation}

In our experiment, we chose the following rotating frame
\begin{equation}
U_{\text{rot}}=e^{i\int_{0}^{t}[H_0-X_3(\tau)\sigma_z \otimes  \mathbf{I}-Y_0(\tau) \mathbf{I}\otimes\sigma_z-Y_3(\tau)\sigma_z\otimes\sigma_z]d\tau}. 
\end{equation}
After dumping the fast oscillating terms according to the rotating wave approximation, one can transform the total Hamiltonian $ H_{\rm total}=H_0+H_{\text{MW}}$ into the effective Hamiltonian
\begin{equation}
\begin{split}
H_{\text{eff}}=&\quad \pi\Omega_1(t)\cos(\phi_1)\sigma_x\otimes |\uparrow\rangle_n\langle\uparrow|_n+\pi\Omega_1(t)\sin(-\phi_1)\sigma_y\otimes |\uparrow\rangle_n\langle\uparrow|_n\\
&+\pi\Omega_2(t)\cos(\phi_2)\sigma_x\otimes |\downarrow\rangle_n\langle\downarrow|_n+\pi\Omega_2(t)\sin(-\phi_2)\sigma_y\otimes |\downarrow\rangle_n\langle\downarrow|_n\\
&+X_3(t)\sigma_z \otimes \mathbf{I}+Y_0(t)\mathbf{I}\otimes\sigma_z+Y_3(t)\sigma_z\otimes\sigma_z. 
\label{RWAEq}
\end{split}
\end{equation}

Now we expand the dilated Hamiltonian $H_{e,n}$ in Eq. \ref{DH} in terms of Pauli operators and the identity matrix $\{\mathbf{I},  \sigma_x,\sigma_y,\sigma_z\}^{\otimes2}$, as
\begin{equation}
\begin{split}
H_{\text{e,n}}=&\quad X_0(t)\mathbf{I} \otimes \mathbf{I} +X_1(t)\sigma_x\otimes \mathbf{I}+X_2(t)\sigma_y\otimes \mathbf{I}+X_3(t)\sigma_z\otimes \mathbf{I} \\ 
&+Y_0(t)\mathbf{I}\otimes\sigma_z+Y_1(t)\sigma_x\otimes\sigma_z+Y_2(t)\sigma_y\otimes\sigma_z+Y_3(t)\sigma_z\otimes\sigma_z,
\end{split}
\label{hsa}
\end{equation}
where $X_i(t)$ and $Y_i (t)$ ($i=0,1,2,3$) are time-dependent real parameters. 

Eq.~\ref{hsa} can be rewritten as
\begin{equation}
\begin{split}
H_{\text{e,n}}=
&\quad[X_1(t)+Y_1(t)]\sigma_x\otimes |\uparrow\rangle_n\langle\uparrow|_n+[X_2(t)+Y_2(t)]\sigma_y\otimes |\uparrow\rangle_n\langle\uparrow|_n\\
&+[X_1(t)-Y_1(t)]\sigma_x\otimes |\downarrow\rangle_n\langle\downarrow|_n+[X_2(t)-Y_2(t)]\sigma_y\otimes |\downarrow\rangle_n\langle\downarrow|_n\\
&+X_0(t)\mathbf{I} \otimes \mathbf{I}+X_3(t)\sigma_z\otimes \mathbf{I} +Y_0(t)\mathbf{I}\otimes\sigma_z+Y_3(t)\sigma_z\otimes\sigma_z.
\end{split}
\label{hsa1}
\end{equation}

By comparing Eq.~\ref{hsa1} with Eq.~\ref{RWAEq}, we thus obtain the following algebraic relations between the tunable MV parameters $\{ \Omega_i,\omega_i, \phi_i| i=1,2\}$ and the desired Hamiltonian parameters $\{ X_i,Y_i| i=0,1,2,3\}$.  Concretely, for the MW amplitudes $\Omega_{1,2}$, we have
\begin{equation}
\begin{split}
\Omega_1(t)=\frac{1}{\pi}\sqrt{(X_1(t)+Y_1(t))^2+(X_2(t)+Y_2(t))^2},\quad 
\Omega_2(t)=\frac{1}{\pi}\sqrt{(X_1(t)-Y_1(t))^2+(X_2(t)-Y_2(t))^2}.
\end{split}
\end{equation}
For the MW frequencies $\omega_{1,2}$, we have
\begin{equation}
\begin{split}
\omega_1(t)=\omega_{\uparrow}+2*(X_3(t)+Y_3(t)),\quad
\omega_2(t)=\omega_{\downarrow}+2*(X_3(t)-Y_3(t)).
\end{split}
\end{equation}
For the MW phases $\phi_{1,2}$, we have
\begin{equation}
\begin{split}
\phi_1(t)=-{\rm arg}[(X_1(t)+Y_1(t))+(X_2(t)+Y_2(t))i],\quad
\phi_2(t)=-{\rm arg}[(X_1(t)-Y_1(t))+(X_2(t)-Y_2(t))i].
\end{split}
\end{equation}

\begin{figure}[h]
	\centering
	\includegraphics[width=1\linewidth] {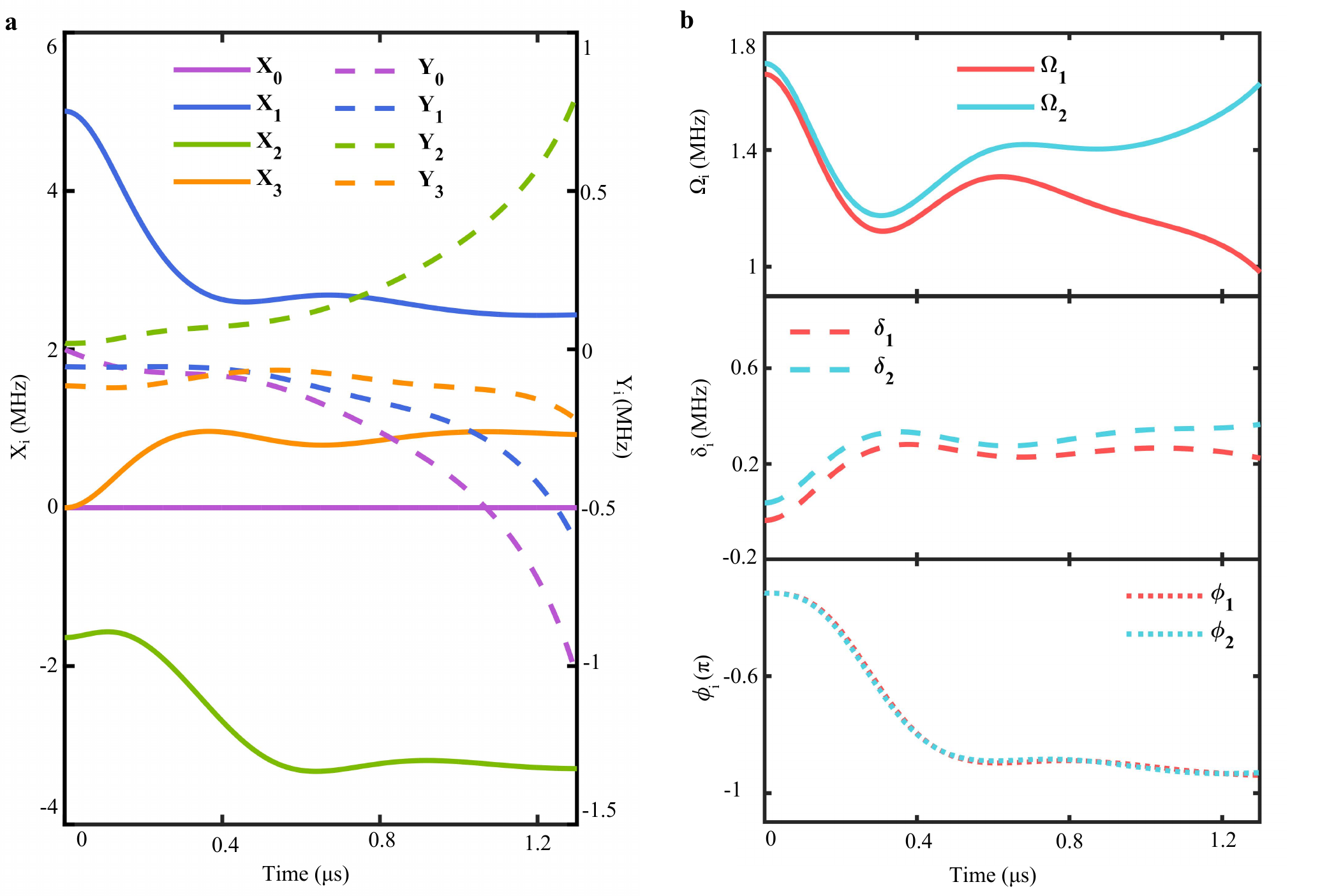}
	\caption{\textbf{Parameters as functions of time.}
\textbf{a}, Parameters \bm{$X_i$} and \bm{$Y_i$} in the dilated Hamiltonian.
\textbf{b}, The microwave frequency detuning, strength and phase. Here, the detuning means the difference between the MW frequency and the resonant frequency, i.e., $\delta_1(t)=\omega_1(t)-\omega_{\uparrow}$ and $\delta_2(t)=\omega_2(t)-\omega_{\downarrow}$.
	}
	\label{exp_parameter}
\end{figure}

To realize the target non-Hermitian twister Hamiltonian $H_e=H(m_1,m_2,k)$, one can obtain the values of  $\{ X_i,Y_i| i=0,1,2,3\}$ from Eq.~\ref{DH}.  As an example,  we show the time dependence of $X_i(t)$ and $Y_i (t)$ in Fig. \ref{exp_parameter}a, with the parameters in $H_{e,n}$ set as $\eta(0)=30, m_{1}=0.9855, m_{2}=0.6$, and $k=0.125\pi$,  $\gamma=3.5$ \footnote{To speedup the evolution process, we product the target non-Hermitian Hamiltonian by  the parameter $\gamma$, i.e., $H_e\rightarrow \gamma H_e$,  so that we can finish the experiment within the coherence time.}.  Fig. \ref{exp_parameter}b shows the time-dependence of the MW detuning ($\delta_i$), amplitude ($\Omega_i$), and phase ($\phi_i$), $i=1,2$.

\subsection{Quantum state tomography}
\begin{figure}[h]
	\centering
	\includegraphics[width=0.9\linewidth] {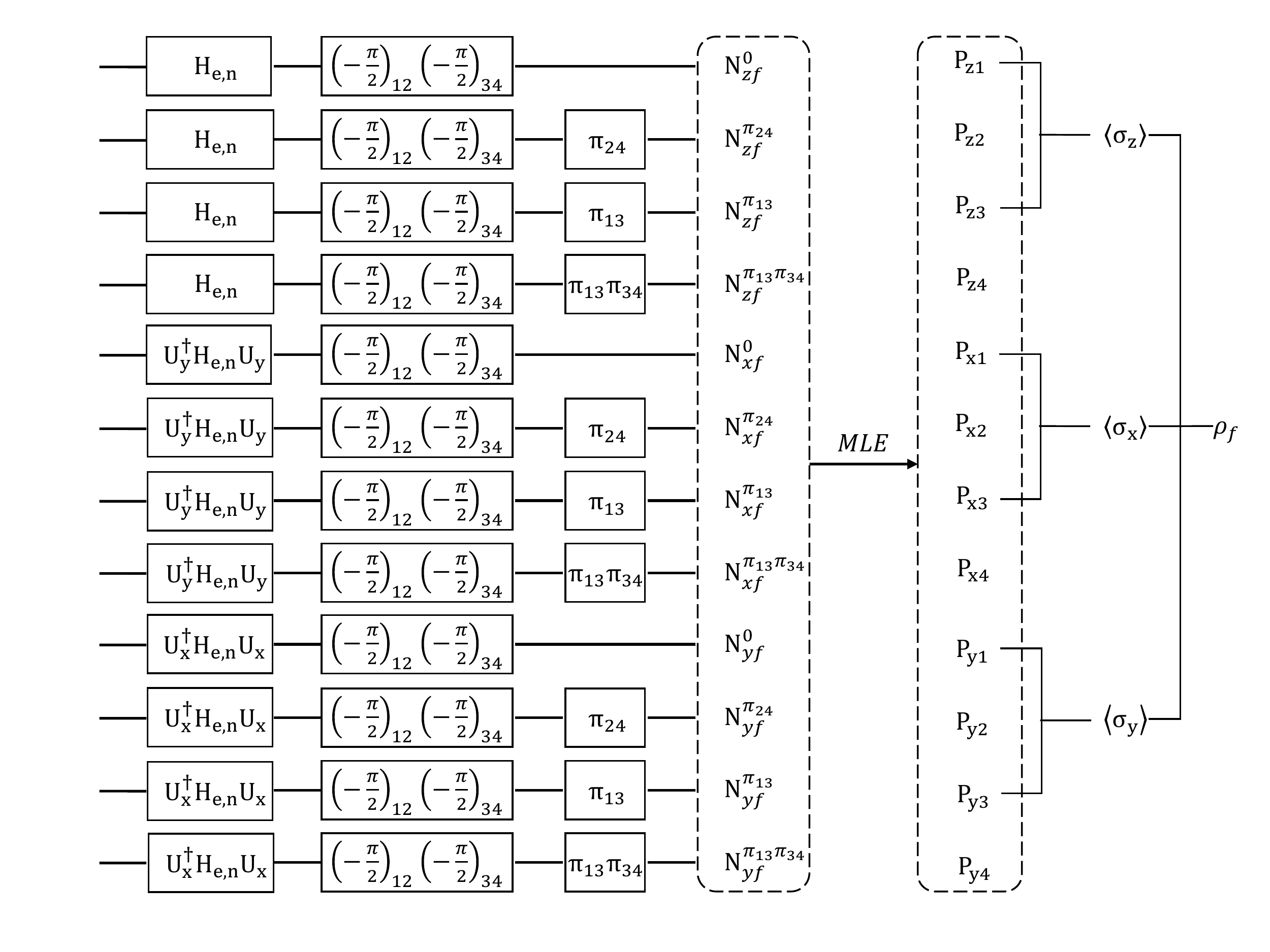}
	\caption{\textbf{Schematic state measurement sequences.}
	 Here $N_{zf}^0$, $N_{zf}^{\pi_{24}}$, $N_{zf}^{\pi_{13}}$ and $N_{zf}^{\pi_{13}\pi_{34}}$ denote the PL rates for measurement of $\langle \sigma_z\rangle$. Same for $\langle \sigma_x\rangle$ and $\langle \sigma_y\rangle$ measurement. $\rho_{f}$ indicates the final state after the the RF $\frac{\pi}{2}$ pulses that are denoted by $(-\frac{\pi}{2})_{12}$ and $(-\frac{\pi}{2})_{34}$.
	}
	\label{Schematic}
\end{figure}

In this subsection, we introduce the method utilized for reconstructing the experimental density matrix of the target qubit. Fig.~\ref{Schematic} shows the schematic sequences. 

Owing to the existence of the ISC \cite{Goldman2015State} and the ESLAC \cite{Smeltzer2009Robust}, the states $|0\rangle_e|\uparrow\rangle_n$, $|0\rangle_e|\downarrow\rangle_n$, $|-1\rangle_e|\uparrow\rangle_n$ and $|-1\rangle_e|\downarrow\rangle_n$ (labeled with numbers from 1 to 4) give rise to different photoluminescence (PL) rates ($N_i$ , $i=1,2,3,4$). To get the exact values of $N_i$, we initialize the system to be $|0\rangle_e|\uparrow\rangle_n$ state through optical pumping and flip the population onto different states with selective MW and RF pulses. For long time measurements, the PL rates sometimes change slightly due to the fluctuation of environment and the contamination of sample. Thus we calibrate the PL rates every couple of days.

After the time evolution process and a subsequent nuclear-spin $\pi/2$ rotation, the final state of the system becomes $|\Psi(t)\rangle = |\psi(t)\rangle_e|\uparrow\rangle_n+\eta(t)|\psi(t)\rangle_e |\downarrow\rangle_n$ in which $|\psi(t)\rangle_e$ is the state of our target qubit. The PL rate of the final state is 
\begin{equation}
N_{zf}^0=\sum_i P_{zi} N_i.
\end{equation}
where $P_{zi}$ ($i=1,2,3,4$) indicates the corresponding populations. 

Now we have obtained the values of $N_{1,2,3,4}$.  Then by flipping the populations between different states and measuring the resultant PL rates, we can solve the population distribution of the final state from the following linear equations:
\begin{equation}
\begin{bmatrix}
N_1&N_2&N_3&N_4\\
N_1&N_4&N_3&N_2\\
N_3&N_2&N_1&N_4\\
N_4&N_2&N_1&N_3
\end{bmatrix}
\begin{bmatrix}
P_{z1}\\P_{z2}\\P_{z3}\\P_{z4}
\end{bmatrix}
=
\begin{bmatrix}
N_{zf}^0\\N_{zf}^{\pi_{24}}\\N_{zf}^{\pi_{13}}\\N_{zf}^{\pi_{13}\pi_{34}}
\end{bmatrix}.
\label{population}
\end{equation}
Here, $\pi_{ij}$ represents the $\pi$-pulse between state $i$ and state$j$; $N_{zf}^{\pi_{13}\pi_{34}}$ denotes the PL rate that is measured after applying $\pi_{34}$ and $\pi_{13}$ sequentially.

As for the measurement of the expectation value of $\sigma_x$ ($\sigma_y$), an $\pi/2$ rotation of the electron spin along the x-axis (y-axis) is required. The time duration of the $\pi/2$ MW pulse should be on the order of 1 $\mu s$ to ensure that it is weak enough to avoid off-resonance driving. Adding this rotation to the experiment circuit will make the total processing time close to the coherence time of our sample. In this case, the fidelity of the final state will decrease due to the gate and decoherence errors. Therefore, we use an alternative approach which is more appropriate for our experiments. Instead of measuring the $\langle \sigma_x\rangle$ ($\langle \sigma_y\rangle$) directly, we rotate the whole system with  $U_{y}=\frac{1}{\sqrt{2}}\begin{pmatrix} 1&-1 \\1&1 \end{pmatrix}$ ($U_{x}=\frac{1}{\sqrt{2}}\begin{pmatrix} 1&i \\i&1 \end{pmatrix}$) and then measure $\langle \sigma_z\rangle$. It can be proved that the two approaches are equivalent. Let $|\psi\rangle$ be one of the eigenvector of Hamiltonian $H_e$, such that $H_e|\psi\rangle=\lambda|\psi\rangle$.
Then we have
\begin{eqnarray}
\langle \psi|\sigma_x|\psi\rangle=\langle \psi|UU^\dagger\sigma_xUU^\dagger|\psi\rangle=\langle \widetilde{\psi}|\sigma_z|\widetilde{\psi}\rangle,
\end{eqnarray}
where $|\widetilde{\psi}\rangle=U^\dagger|\psi\rangle$ is the eigenvector of $\widetilde{H_e}$ with the same eigenvalue $\lambda$ according to the following equation
\begin{equation}
\widetilde{H_e}|\widetilde{\psi}\rangle=U^\dagger H_e UU^\dagger|\psi\rangle=U^\dagger \lambda|\psi\rangle=\lambda |\widetilde{\psi}\rangle.
\end{equation}

As demonstrated in Fig. \ref{Schematic}, we implement the evolution under $U_{y}^\dagger H_{e,n} U_{y}$ ($U_{x}^\dagger H_{e,n} U_{x}$) and measure the final state in z-basis. In this way, we get another two groups of linear equations which are similar to Eq. \ref{population}.
Notice that solving $P_{xi}$, $P_{yi}$ and $P_{zi}$ ($i=1,2,3,4$) separately sometimes will lead to a result that violates some essential properties of the density matrix of a pure state, such as having a value of $tr \rho_{f}^2$ that is not equal to 1. To avoid this problem, here we use the maximum likelihood estimation (MLE) \cite{Hayashi2005Asymptotic} method to solve the populations all together under the following constraints:
\begin{equation}
\left\{\begin{aligned}
&P_{z1}+P_{z2}+P_{z3}+P_{z4}=1,\\
&P_{x1}+P_{x2}+P_{x3}+P_{x4}=1,\\
&P_{y1}+P_{y2}+P_{y3}+P_{y4}=1,\\
&\left(\frac{P_{z1}-P_{z3}}{P_{z1}+P_{z3}}\right)^2+\left(\frac{P_{x1}-P_{x3}}{P_{x1}+P_{x3}}\right)^2+\left(\frac{P_{y1}-P_{y3}}{P_{y1}+P_{y3}}\right)^2=1.\\
\end{aligned}
\right.
\end{equation}

Knowing $P_{xi}$, $P_{xi}$ and $P_{zi}$, we can determine the expectation value of $\sigma_x$, $\sigma_y$ and $\sigma_z$ for the state $|\psi(t)\rangle$ by renormalizing the populations in the $|\uparrow\rangle_n$ subspace, i.e. 
$\langle \psi(t)|\sigma_x|\psi(t)\rangle=\frac{P_{x1}-P_{x3}}{P_{x1}+P_{x3}}$, $\langle \psi(t)|\sigma_y|\psi(t)\rangle=\frac{P_{y1}-P_{y3}}{P_{y1}+P_{y3}}$ and $\langle \psi(t)|\sigma_z|\psi(t)\rangle=\frac{P_{z1}-P_{z3}}{P_{z1}+P_{z3}}$.

\begin{figure}[h]
	\centering
	\includegraphics[width=1\linewidth] {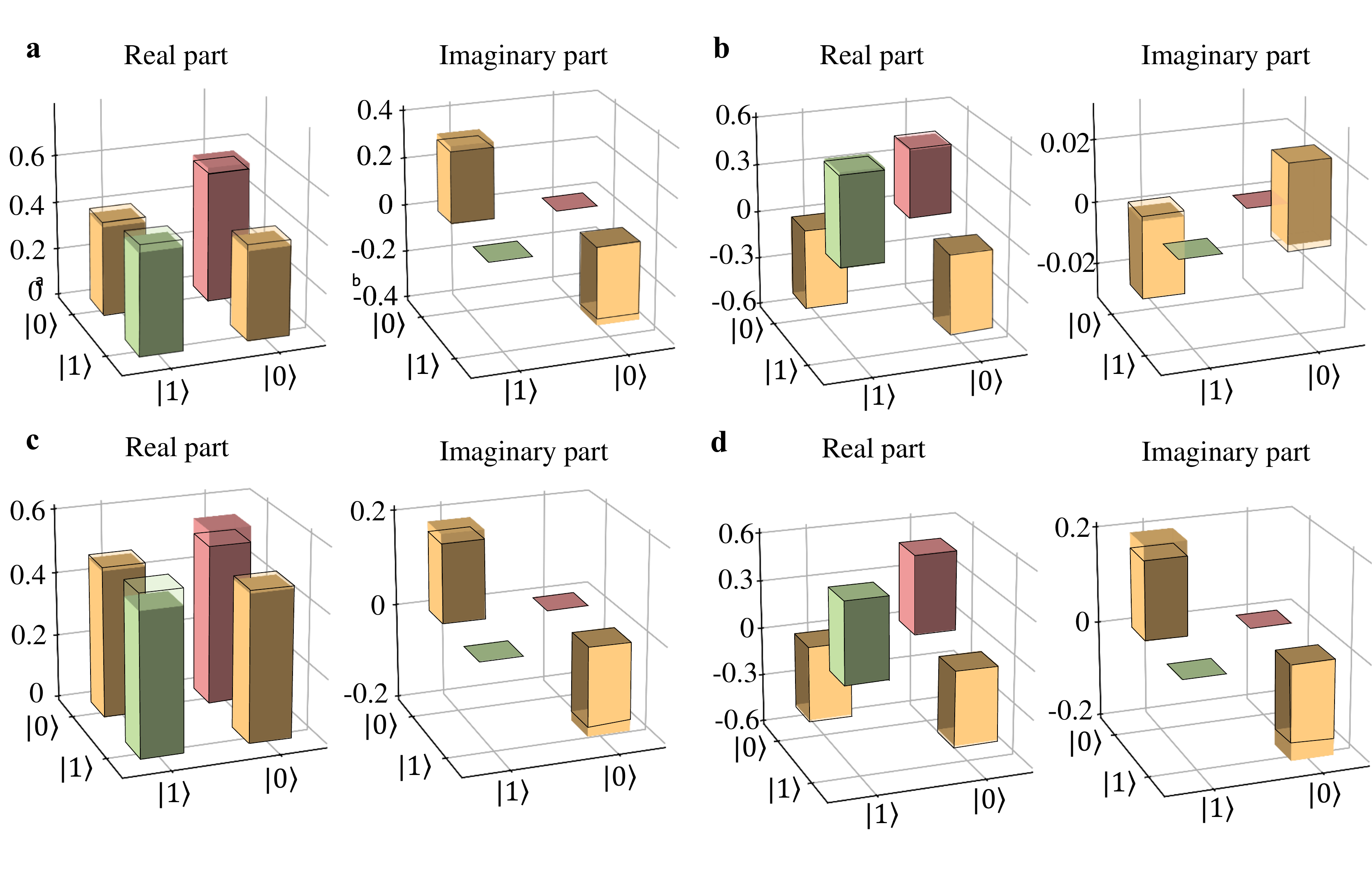}
	\caption{\textbf{The final states reconstructed through quantum state tomography.}
\textbf{a}, The density matrix elements of the final electron spin state measured after evolving under $H(k)$ with the parameters $m_{1}=0.5338$, $m_{2}=0.6$ and $k=0.125\pi$. 
\textbf{b}, The density matrix elements of the final state reached by evolving under $-H(k)$ with the same parameters as those in \textbf{a}. 
\textbf{c}, The density matrix elements of the electron spin final state measured after evolving under $iH(k)$ with the parameters  $m_{1}=0.5338$, $m_{2}=0.6$ and $k=2\pi$. 
\textbf{d}, The density matrix elements of the final state reached by evolving under $-iH(k)$ with the same parameters as those in \textbf{c}. 
The transparent columns in \textbf{a}-\textbf{d} denote the matrix elements of the corresponding ideal final states.}
	\label{Tomography}
\end{figure}

Then the density matrix can be reconstructed by 
\begin{equation}
\rho_{f}=\frac{1}{2}(I+\vec{r}\cdot\vec{\sigma}).
\end{equation}
where $\vec{r}=(\langle \sigma_x\rangle, \langle \sigma_y\rangle, \langle \sigma_z\rangle)$ is the Bloch vector, and $\vec{\sigma}=(\sigma_x, \sigma_y, \sigma_z)$ denotes the vector of Pauli matrices.
Fig. \ref{Tomography} shows some examples of the measured density matrix elements of the final states.

\newpage
\subsection{The non-unitary time evolution}
To provide further evidence on the validity of the dilated Hamiltonian, we measure the non-unitary time evolution of the renormalized
populations:
\begin{eqnarray}
P_{-1}^{z}&=&Tr(\rho_{e}|-1\rangle_{e}\langle-1|)=\frac{P_{z3}}{P_{z1}+P_{z3}},\nonumber\\
P_{-1}^{x}&=&Tr(\rho_{e}|S_{x};-\rangle_{e}\langle S_{x};-|)=\frac{P_{x3}}{P_{x1}+P_{x3}},\label{partequations}\\
P_{-1}^{y}&=&Tr(\rho_{e}|S_{y};-\rangle_{e}\langle S_{y};-|)=\frac{P_{y3}}{P_{y1}+P_{y3}},\nonumber
\end{eqnarray}
where $|S_{x};-\rangle_{e}=\frac{1}{\sqrt{2}}(|0\rangle_{e}-|-1\rangle_{e})$ and $|S_{y};-\rangle_{e}=\frac{1}{\sqrt{2}}(|0\rangle_{e}-i|-1\rangle_{e})$.

\begin{figure}[h]
	\centering
	\includegraphics[width=1\linewidth] {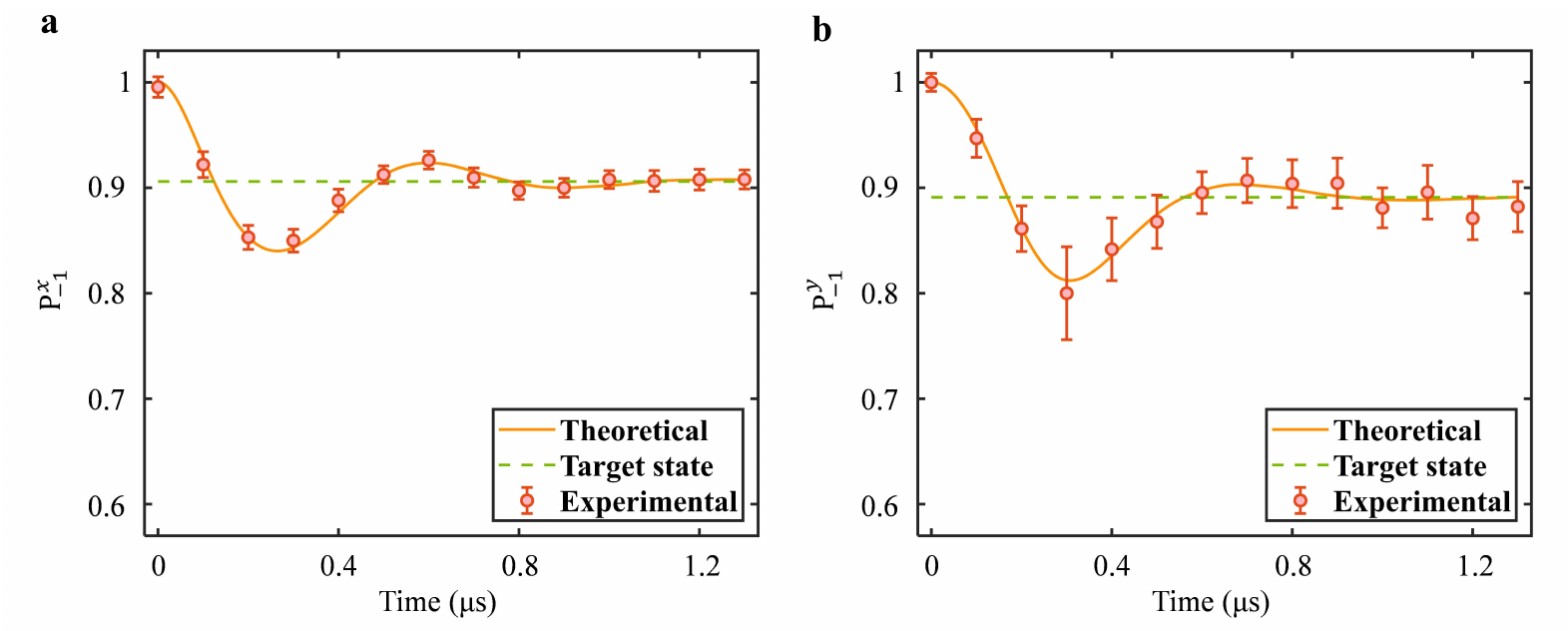}
	\caption{\textbf{The time evolution curves.}
\textbf{a}, The time evolution of the renormalized population {$P_{-1}^{x}$}. 
\textbf{b}, The time evolution of the renormalized population {$P_{-1}^{y}$}. The parameters characterizing the underlying twister Hamiltonian are chosen as $m_{1}=0.4928$, $m_{2}=0.6$, $k=1.875\pi$ and $\gamma=3.5$ for \textbf{a}, whereas $m_{1}=0.9855$, $m_{2}=0.6$, $k=0.125\pi$ and $\gamma=3.5$ for \textbf{b}.
	}
	\label{TimeEvolution}
\end{figure}

An example of the time evolution of $P_{-1}^{z}$ has been presented in Fig. 2b in the main text. For completeness, here we show the time evolution of $P_{-1}^{x}$ and $P_{-1}^{y}$. The experimental results (dots with error bars) shown in Fig. \ref{TimeEvolution} are in good agreement with the theoretical curves (solid lines) predicted by numerical simulations. We see that the state decays to the desired eigenstate (dashed lines) of the twister Hamiltonian after long time evolution.

\clearpage
\newpage

\section{Experimental observations associated with the knotted topological phases}
\subsection{Trajectories of eigenstates on the Bloch sphere}

The two-band 1D non-Hermitian twister Hamiltonian $H(k)$ has three distinct topological phases: the Hopf link phase, the unknot phase and the unlink phase (see Fig. 1b in the main text). Here we probe the knot topology by preparing the eigenstates of $H(k)$ with different parameter sets ($m_{1},m_{2}$). As the momentum $k$ sweeps from 0 to $2\pi$, the eigenstates form curves on the Bloch sphere. 

In simulating the Hermitian topological phases in Bloch space, the ground states with different momentums can be prepared with the adiabatic passage approach. However, the adiabatic passage method for the Hermitian models cannot straightforwardly carry over to the non-Hermitian scenario with complex eigenvalues. To prepare the eigenstates of the twister model, here we utilize the feature of non-unitary evolution governed by the non-Hermitian Hamiltonian. Specifically, suppose the target system is initially at the state $|\psi(0)\rangle=c_1|R_1\rangle+c_2|R_2\rangle$, where $|R_{1,2}\rangle$ are the right eigenvectors of the twister Hamiltonian with eigenvalues $\lambda_{1,2}$. Suppose $\text{Im}(\lambda_1)>\text{Im}(\lambda_2)$, then the system would decay to  $|R_{1}\rangle$ in the long time limit. Hence, one can prepare the eigenstate $|R_1\rangle$ of the twister Hamiltonian based on the long time evolution of the system. Similarly, one can prepare the other  eigenstate $|R_{2}\rangle$ by simply changing $H$ into $-H$. 
We remark that such progress  relies crucially on the imaginary part of the eigenvalues and is suitable for most cases of the parameters $(m_1,m_2,k)$. For those special cases with real eigenvalues ($k=\pi, 2\pi$ for the Hopf link phase; $k=2\pi$ for the unknot phase), one can prepare the corresponding eigenstates through the dynamical evolution of $\pm iH$.

\begin{figure}[h]
	\centering
	\includegraphics[width=1\linewidth] {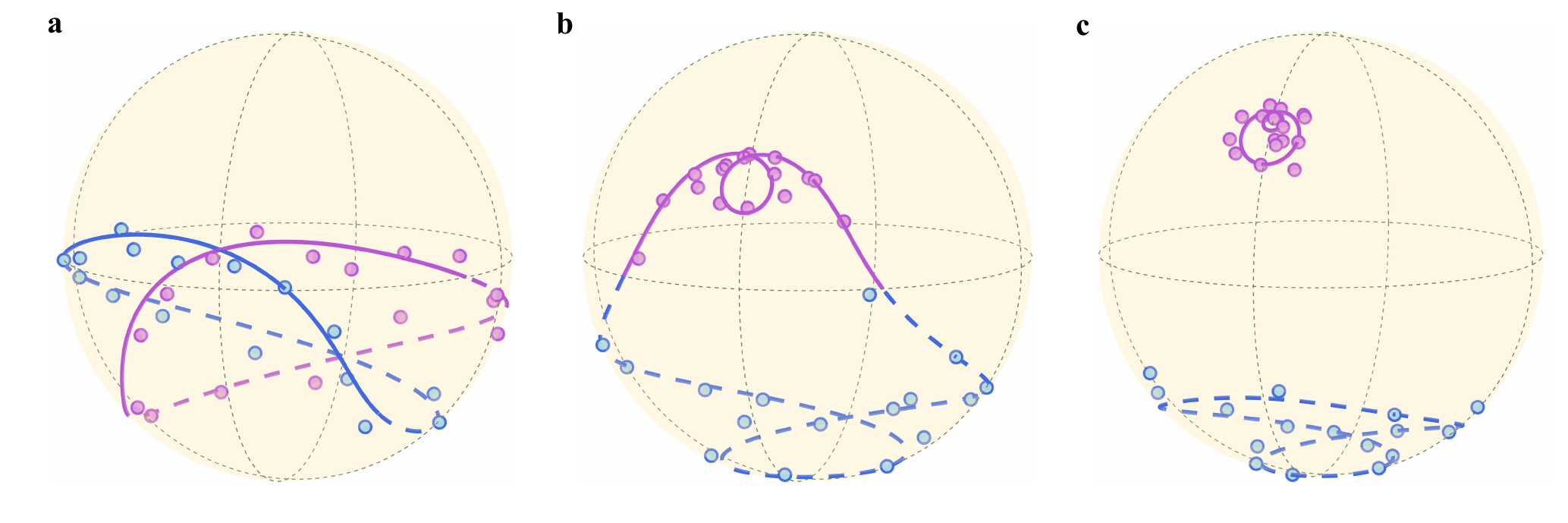}
	\caption{\textbf{Trajectories of the eigenvectors on the Bloch sphere.} Parameters: \textbf{a}, $m_{1}=0.5338$, $m_{2}=0.6$ (the Hopf link phase); \textbf{b}, $m_{1}=1.2730$, $m_{2}=0.6$ (the unknot phase); and \textbf{c}, $m_{1}=1.8889$, $m_{2}=0.6$ (the unlink phase). The purple and blue curves denote the theoretical trajectories of the two eigenvectors $|R_{1}(k)\rangle$ and $|R_{2}(k)\rangle$ as k spans from 0 to $2\pi$. The solid and dashed curves are achieved based on the evolution of $H(k)$ and $-H(k)$ respectively. The purple and blue dots represent the experimental results.
	}
	\label{Signature}
\end{figure}

Fig. \ref{Signature} shows the three different trajectories of the eigenvectors on the Bloch sphere, with each trajectory corresponding to one topological phase.
It is evident that different knot structures can be identified by the distinctive crossing features of the eigenvector loops. As for the Hopf link phase, each of the two eigenstates $|R_{1,2}(k)\rangle$ gives rise to a closed loop. The two closed loops intersect with each other twice. For the unlink phase, the two eigenstates also form two separate closed loops but with no intersection point. In the case of the unknot phase, each eigenstate gives an open curve. The two open curves meet at the points of $k=2\pi$, making single closed loop on the Bloch sphere. The corresponding experimental data and the error bars obtained by Monte Carlo simulation are listed in Tables \ref{table1}-\ref{table6}.

\newpage
\subsection{Experimental knot band structures}

Generally, the N dimensional non-Hermitian Hamiltonian $H$ takes the following eigenspectra formula
\begin{equation}
    H=\sum_{i=1}^N E_i |R_i\rangle\langle L_i|,
\end{equation}
where $E_i$'s are the eigenvalues of the Hamiltonian and can be complex,  $|R_i\rangle$ and $|L_i\rangle$ denote the right  eigenvectors of $H$ and $H^{\dag}$ , respectively, i.e., 
\begin{equation}
H|R_{i}\rangle=E_{i}|R_{i}\rangle,\qquad H^{\dag}|L_{i}\rangle=E_{i}^{*}|L_{i}\rangle.
\end{equation}
And the eigenvectors obey the following biorthogonal normalization conditions,
\begin{equation}
\langle L_{m}|R_{n}\rangle=\delta_{mn}, \quad  \forall m,n\,\in [1,N].
\label{orthonormal}
\end{equation}
With the above relation,  one can express the eigenenergies of the non-Hermitian Hamiltonian as follows
\begin{equation}
E_{n}(k)=\langle L_{n}(k)|H(k)|R_{n}(k)\rangle.
\end{equation}
In the special case of Hermitian $H=H^\dag$, the eigenvalues $\{E_i\}$ become real, and the corresponding eigenstates $|R_i\rangle=|L_i\rangle$.


Here we focus on the two-band non-Hermitian twister Hamiltonian $H(k)$ with $k$ in the first Brillouin zone, which can be formally  expressed as 
\begin{equation}
H(k)=E_{1}(k)|R_{1}(k)\rangle\langle L_{1}(k)|+E_{2}(k)|R_{2}(k)\rangle\langle L_{2}(k)|.
\end{equation}
With the biorthogonal relation in Eq.~\ref{orthonormal}, one can express the eigenenergies of the non-Hermitian twister Hamiltonian by
\begin{equation}
E_{n}(k)=\langle L_{n}(k)|H(k)|R_{n}(k)\rangle, \quad (n=1,2).
\end{equation}

\begin{figure}[h]
	\centering
	\includegraphics[width=1\linewidth] {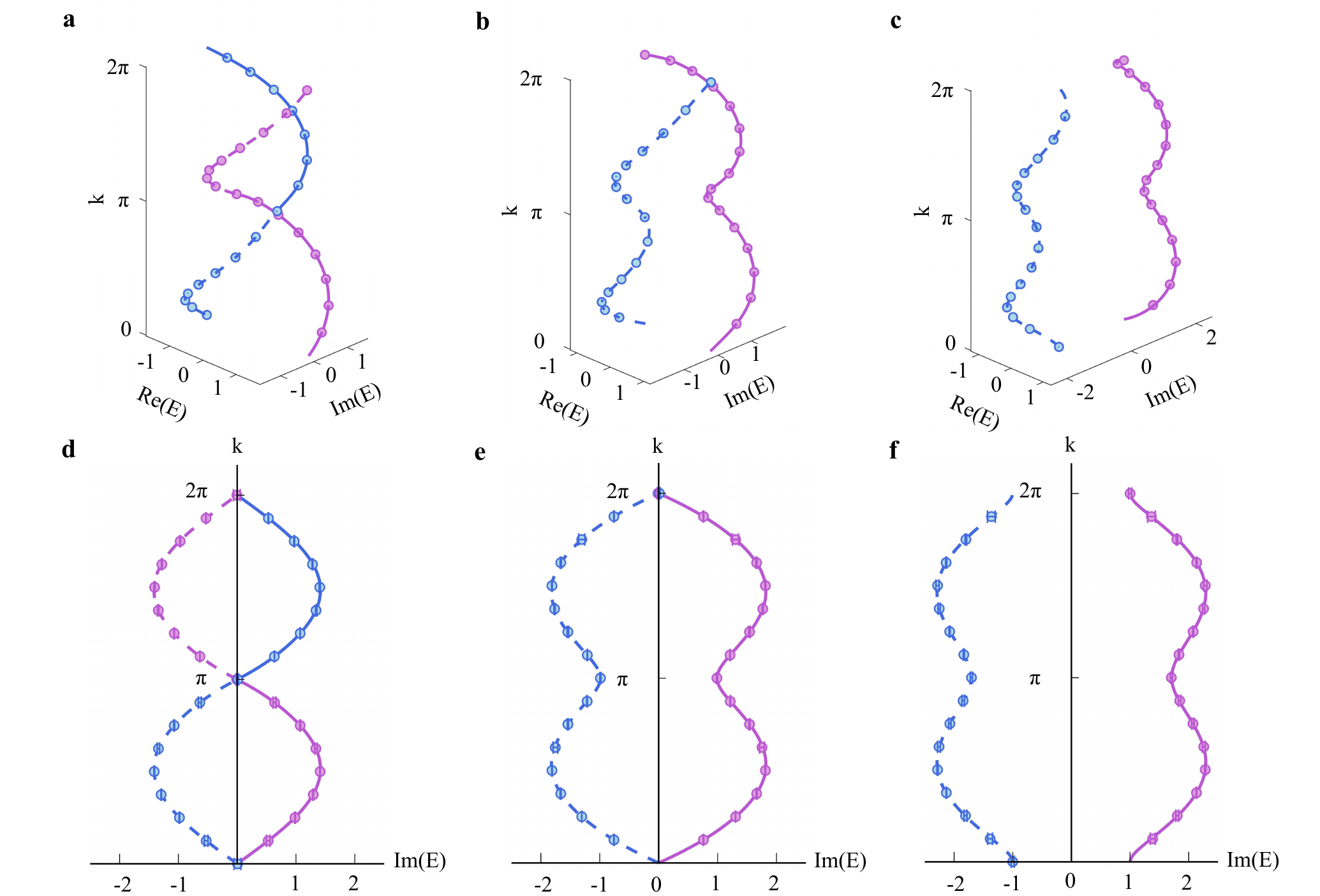}
	\caption{\textbf{Non-Hermitian band structures of the twister Hamiltonian.}
Parameters: \textbf{a}, $m_{1}=0.5338$, $m_{2}=0.6$ (the Hopf link phase); \textbf{b}, $m_{1}=1.2730$, $m_{2}=0.6$ (the unknot phase); and \textbf{c}, $m_{1}=1.8889$, $m_{2}=0.6$ (the unlink phase).
\textbf{a-c}, Braiding patterns of the complex eigenenergy bands. 
\textbf{d-f}, Projection of the 3D complex eigenenergy bands onto the 2D (\bm{$\textrm{Im}(E), k$}) plane.
The purple and blue curves denote the theoretical predictions of the two  eigenenergy strings. The solid (dashed) curves denote the non-Hermitian bands with positive (negative) imaginary part of eigenenergy. The corresponding experimental results are represented by the purple and blue dots with error bars. 
	}
	\label{BandStructure}
\end{figure}

In our experiment, we  implement the right eigenstates $|R_{1,2}(k)\rangle$ based on the non-unitary dynamical evolution of electron spin system. The left eigenstates $|L_{n}(k)\rangle$ are calculated  based on the biorthogonal normalization conditions in Eq.~\ref{orthonormal}. Then we can obtain the band structures of the twister Hamiltonian in the first Brillouin zone based on the experimental data. 
Here we plot the braided band structures of the twister Hamiltonian in Fig. \ref{BandStructure}, together with their 2D projections on the plane spanned by the imaginary part of energy $Im(E)$ and the momentum $k$. The knot topology is apparent from the braiding patterns of the two eigenenergy strings. In addition, different knots give rise to different projections which can be characterized by the overlapping points. For the Hopf link phase, projections of the two bands intersect at $(0, 0)$ and $(0, \pi)$. Each band has eigenenergies with positive and negative imaginary parts. For the unknot phase, the projection has one overlapping point at $(0, 0)$. Signs of $Im(E)$ remain unchanged between $k=0$ and $k=2\pi$. As for the unlink phase, $Im(E)$ does not vanish for any $k$. Therefore, no overlapping point can be seen in the projection. We see from Fig.\ref{BandStructure} that the experimental results (colored dots) match precisely with the theoretical trajectories (solid and dashed lines) of the bands.



\subsection{Calculation of the topological invariant Q with experimental data }
In this subsection, we calculate the topological invariant directly from our experimental data. 

The global biorthogonal Berry phase is defined as \cite{Liang2013Topological}
\begin{equation}
Q=\frac{1}{\pi}\oint dkTrC_{k}, 
\label{DefineQ}
\end{equation}
where $C_{k}$ is the non-Abelian Berry connection. For the two-band non-Hermitian Hamiltonian, 
\begin{equation}
C_{k}=i
\begin{bmatrix}
\langle L_{1}(k)|\partial_{k}|R_{1}(k)\rangle&\langle L_{1}(k)|\partial_{k}|R_{2}(k)\rangle\\
\langle L_{2}(k)|\partial_{k}|R_{1}(k)\rangle&\langle L_{2}(k)|\partial_{k}|R_{2}(k)\rangle
\end{bmatrix}.
\end{equation}
To calculate $Q$ with our discretized experimental data, we use the method proposed in Ref. \cite{fukui2005chern}. 
\begin{equation}
\left\{\begin{aligned}
D_{n}^{i}=\ln \frac{\langle L_{n}(k_{i+1})|R_{n}(k_{i})\rangle}{\langle L_{n}(k_{i})|R_{n}(k_{i})\rangle}, \\
w=\frac{1}{\pi}\Sigma_{i,n}  \text{Im}(D_{n}^{i}).
\end{aligned}
\right. 
\label{DiscreteTopIndex}
\end{equation} 
In Table \ref{Q}, we show the global biorthogonal Berry phase extracted from the experimental and numerically simulated data with the same values of parameter $m_{1}$ as in Fig.~\ref{Signature}.

\begin{table*}[!h]
	\centering
	
	\begin{tabular}{c|c|c}
		\hline
		Parameter $m_{1}$ & Experimental $Q$ & Theoretical $Q$ \\
		\hline
		0.5338 ($k:0\rightarrow2\pi$)&$2.0193 \pm 0.0262$&$2.0000$\\
		1.2730 ($k:0\rightarrow4\pi$)&$2.0105 \pm 0.0251$&$2.0000$\\
		1.8889 ($k:0\rightarrow2\pi$)&$0.0116 \pm 0.0296$&$0.0000$\\
		\hline
	\end{tabular}
	\caption{\textbf{The global biorthogonal Berry phase calculated from the experimental data and their theoretical predictions.} The second column shows the results calculated with our experimental data. Here, the uncertainty is calculated with the Monte-Carlo method. The third column shows the simulated result by using discretized momentum points in the Brillouin zone (BZ).}
	\label{Q}
\end{table*}
For the cases of $m_{1}=0.5338$ (the Hopf link phase) and $m_{1}=1.8889$ (the unlink phase), the eigenstates are 2$\pi$ periodic for the momentum $k$. Thus the integration in Eq. \ref{DefineQ} is performed over a region with k from 0 to 2$\pi$. As shown in Table \ref{Q}, the topological invariants Q extracted from experimental data are 2.0193(262) and 0.0116(296) for the cases of $m_{1}=0.5338$  and $m_{1}=1.8889$, respectively.
As for the $m_{1}=1.2730$ case (the unknot phase), the eigenstates are 4$\pi$ periodic for the momentum $k$. Therefore, to perform the closed loop integral,  $k$ should sweep from 0 to 4$\pi$. Due to the fact that $H(k)=H(k+2\pi)$, evolving under $H(k)$ with $k$ goes from 2$\pi$ to 4$\pi$ will only lead to the same result as $k$ lies between 0 and 2$\pi$. Thus we obtain the closed trajectory of one band by evolving under $H(k)$ for $k=[0, 2\pi]$ and $-H(k)$ for $k=[2\pi, 4\pi]$. Likewise, the other band is prepared by evolving under $-H(k)$ for $k=[0, 2\pi]$ and $H(k)$ for $k=[2\pi, 4\pi]$. We find that $Q$ calculated with numerically simulated data is equal to 2 when we integrate from $k=0$ to $k=4\pi$ and conclude that $Q=1$ if $k$ only sweeps from 0 to 2$\pi$, which is in agreement with Ref. \cite{Hu2021Knots}.

\subsection{Topological phase transition at exceptional points}
\begin{figure}[h]
	\centering
	\includegraphics[width=0.88\linewidth] {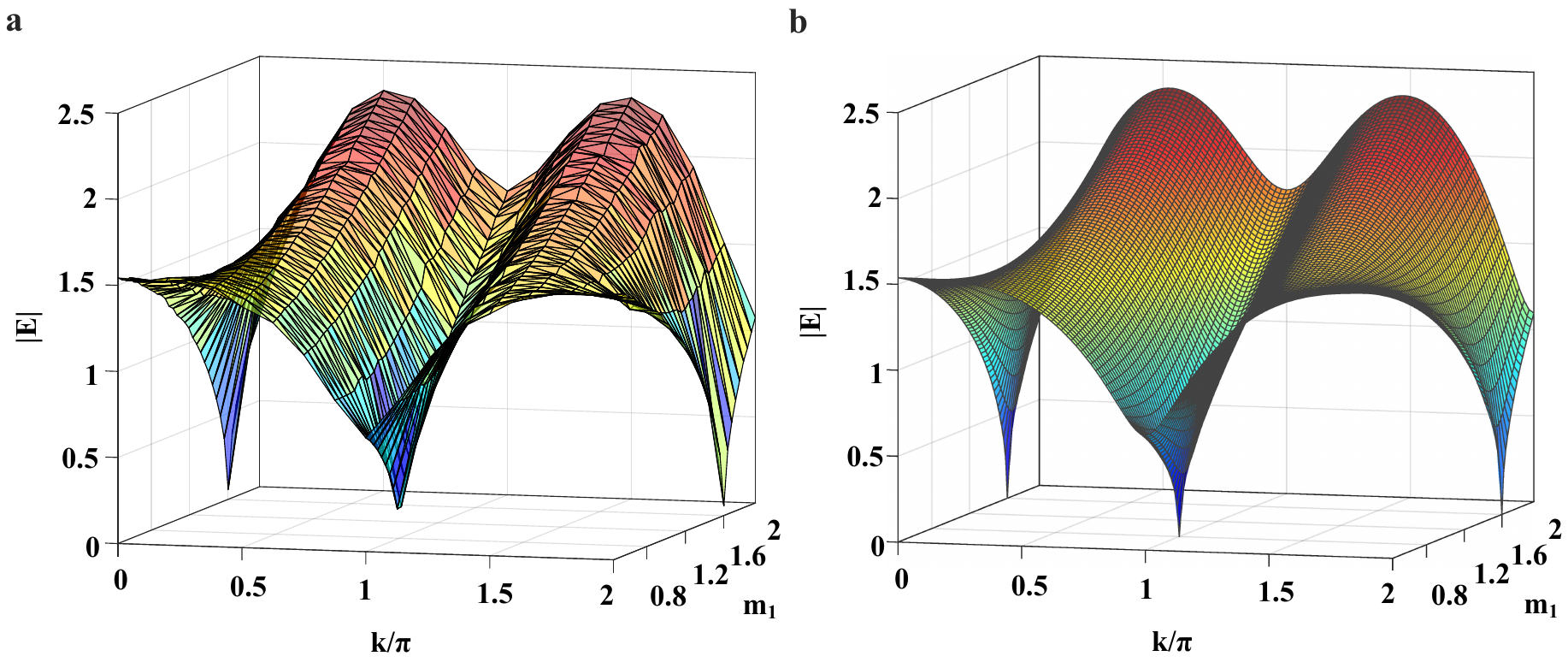}
	\caption{\textbf{Topological phase transitions at EPs.}
Parameters: $m_{1}=0.4106+l*4/\pi^{4}, l\in [1,3]$; $m_{2}=0.6$; $k=n\pi/8, n\in [1,16]$. Energies are measured on a $37\times16$ grid in the ($m_{1},k$) plane.
\textbf{a}, Experimentally measured eigenenergies of the twister Hamiltonian. Two EPs involved in the phase transitions can be observed at ($0.8,\pi$) and ($1.6,2\pi$). 
\textbf{b}, The corresponding theoretical energy dispersion. 
	}
	\label{EnergyMesh}
\end{figure}

Transitions between different knots take place at band degeneracy points where the two eigenenergy strings intersect. For the parametric twister Hamiltonian with $m_{2}$ fixed to 0.6 in our experiment, the three different phases are separated by a kind of band degeneracy called the exceptional point (EP). At EPs, both the eigenvalues and eigenstates of the non-Hermitian Hamiltonian coalesce \cite{Heiss2012The}. Here we demonstrate the phase transitions at EPs in Fig. \ref{EnergyMesh}. As can be seen, the transition between the Hopf link phase and the unknot phase occurs through an EP located at $(m_{1},k)=(0.8,\pi)$. The transition between the unknot phase and the unlink phase is accompanied by an EP located at $(m_{1},k)=(1.6,2\pi)$. 

\subsection{Details on state fidelity}
\begin{figure}[h]
	\centering
	\includegraphics[width=0.84\linewidth] {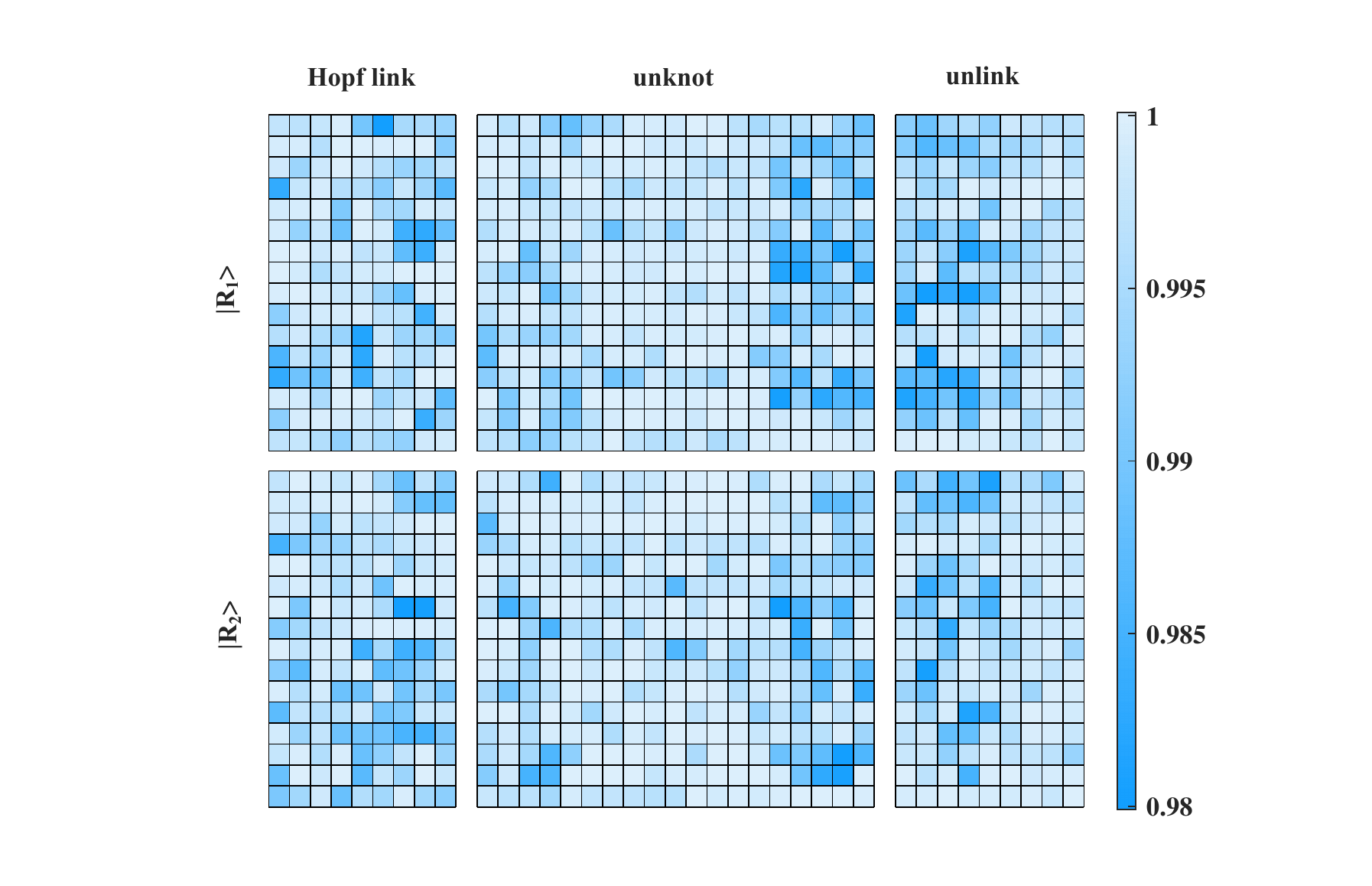}
	\caption{\textbf{State fidelities for different parameter \bm{$m_{1}$} and momenta \bm{$k$}.}
Parameters: $m_{1}=0.4106+l*4/\pi^{4}, l\in [1,3]$; $m_{2}=0.6$; $k=n\pi/8, n\in [1,16]$. Energies are measured on a $37\times16$ grid in the ($m_{1},k$) plane.
The left, the middle and the right panels have the parameter $m_{1}$s belong to different topological phases. The upper and lower panels correspond to the two non-Hermitian bands. The average fidelities are $99.7(3)\%$, $99.7(4)\%$ and $99.5(5)\%$ for the Hopf link phase, the unknot phase and the unlink phase, respectively.
	}
	\label{FidelityDistribution}
\end{figure}

State fidelity is calculated by $F_{k}=\langle\psi_{k}|\rho_{k}|\psi_{k}\rangle$, where $\rho_{k}$ is the measured state density matrix and $|\psi_{k}\rangle$ is the eigenstate in theory. In this work, our input data set for unsupervised learning of the NH twister Hamiltonian is experimentally acquired through quantum simulation of $H(k)$ with $m_{2}$ fixed to 0.6 and $m_{1}$ set in the range from 0.4517 to 1.9300. For each ($m_{1}, m_{2}$), $k$ varies from 0 to $2\pi$ with an equal interval of $0.125\pi$. Fig. 2c in the main text shows the pie chart of the state fidelity. Here, we plot a color map in Fig. \ref{FidelityDistribution} to show more detailed distributions of the fidelity.
\newpage

\newpage
\section{ Introduction to diffusion map}
Here we briefly introduce the diffusion map method for unsupervised machine learning. Diffusion map is  a kind of typical manifold learning algorithm \cite{Coifman2005Geometric,Coifman2005Geometrica,Coifman2006Diffusion}, which provides dimensional reduction non-linearly  and cluster unlabelled data samples  without {\it a priori} knowledge, i.e. in an unsupervised manner. 
It combines the heat diffusion with the random walk Markov chain.  Specifically, given a set of input data ${\bf X } = \{{\bf x}^{(1)},\,{\bf x}^{(2)},\,\cdots {\bf x}^{(L)} \}$, where ${\bf x}^{(i)}$ represents the $i$-th data point in complex space $\mathbb{C}^d$.   The connectivity between two points ${\bf x}^{(l)}$ and ${\bf x}^{(l')}$ is described by the local similarity, which is required to be positive definite and symmetric.  For example, one can measure the distance between two samples based on  the Gaussian kernel
\begin{equation}
\mathcal{K}_{l,l'} = \exp\left(-\frac{\|{\bf x}^{(l)}-{\bf x}^{(l')}\|_{\mathbb{L}_p}^2}{2\epsilon}\right),
\end{equation}
where $\|{\bf x}^{(l)}-{\bf x}^{(l')}\|_{\mathbb{L}_p}$ represents the $\mathbb{L}_p$-norm distance between two points ${\bf x}^{(l)}$ and ${\bf x}^{(l')}$, the variance $\epsilon$ is a small quantity to be adjusted. A number of recent works have reported the  applications of  $p=1,\, 2, \infty$ cases in unsupervised clustering topological phases \cite{Rodriguez-Nieva2019Identifying,Scheurer2020Unsupervised,Che2020Topological,Long2020Unsupervised,Yu2021Unsupervised}. When $p=2$, the distance is the familiar Euclidean distance. With such kernel,  one can define the one-step transition matrix $\mathcal{P}$ of Markovian random walk  between two points ${\bf x}^{(l)}$ and ${\bf x}^{(l')}$ in $\mathbb{C}^d$  as follows
\begin{equation}
\mathcal{P}_{l,l'}=\frac{\mathcal{K}_{l,l'}}{\sum_{l'}\mathcal{K}_{l,l'}},
\end{equation}
with $\mathcal{P}_{l,l'}$ obeying the probability conservation condition $\sum_{l}\mathcal{P}_{l,l'}=1$.  Then after $2t$ steps of random walk, the connectivity between the two points ${\bf x}^{(l)}$ and ${\bf x}^{(l')}$ is represented by the diffusion distance
\begin{equation}
\begin{aligned}
D_{t}(l, l')=D_{t}({\bf x}^{(l)}, {\bf x}^{(l')})= \sum_{k=1}^{L}\frac{\left(\mathcal{P}^t_{l,k}-\mathcal{P}^t_{l',k}\right)^2}{\sum_{j}\mathcal{K}_{k,j}}= \sum_{k=1}^{L-1}\lambda_{k}^{2t}[(\psi_k)_l - (\psi_k)_{l'}]^2\geq 0,
\end{aligned}
\end{equation}
with $\{\psi_k\}$ denoting  the set of right eigenvectors of $\mathcal{P}$, $\mathcal{P}\psi_k =\lambda_k \psi_k$, $k=0,1,... L-1$, the corresponding eigenvalues rank in descending order, i.e. $\lambda_0=1\geq \lambda_1\geq\cdots \geq \lambda_{L-1}$.  Here we note that $k=0$ term does not contribute because the corresponding right eigenvector is constant with all vector elements equivalent. 

Based on the  mapping,  
\begin{equation}\label{diff_map_pbc}
{\bf x}^{(l)}\rightarrow \Psi_t^{(l)} : = [\lambda_1^t(\psi_1)_l, \lambda_2^t(\psi_2)_l, \cdots , \lambda_{L-1}^t(\psi_{L-1})_l],
\end{equation}
one can recast the distance between samples ${\bf x}^{(l)}$ and ${\bf x}^{(l')}$  as the Euclidean distance in $\Psi$ space
\begin{equation}
D_{t}({\bf x}^{(l)}, {\bf x}^{(l')}) =\|\Psi_l - \Psi_{l'}\|^2_{\mathbb{L}_2}.
\end{equation}
After $t\rightarrow \infty$ steps, only those few components  with largest $|\lambda_k|\approx 1$ are  prominent, owing to the  term $\lambda_k^t$ in $\Psi_t$. Hence almost all the efficient distance information of the original data samples in set {\bf X} is encoded in such few components. Then the original samples ${\bf x}^{(l)}$ with higher dimension are reduced to the lower ones in Euclidean space, and then one can apply the usual clustering method (e.g. $k$-means) in $\Psi$ space to cluster the corresponding unlabelled samples without {\it a priori} knowledge.  Specifically,  the number of $|\lambda_k|\approx1$ equals to the number of unlabelled topological clusters in clustering the topological phases of quantum models. Hence it is possible for such algorithm to detect unknown topological phases.


\newpage
\section{Classifying non-Hermitian knotted phases based on diffusion map}
Here we provide more details about the unsupervised machine learning non-Hermitian knotted topological phases based on the diffusion map method. 

\subsection{Numerical simulated result }
\begin{figure}[h]
\centering
\includegraphics[width=0.9\linewidth]{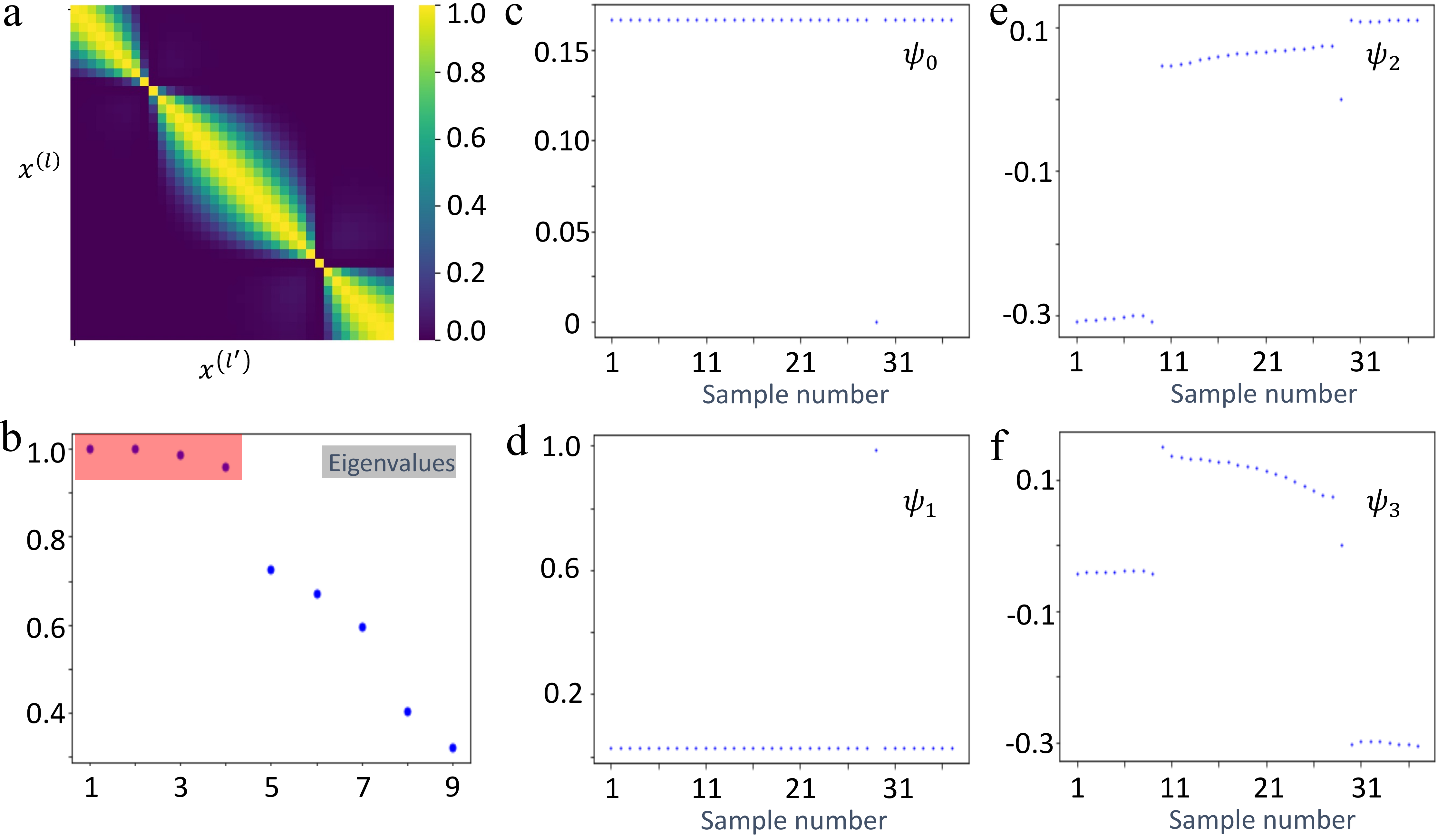}
\caption{{ Classification of 1D Non-Hermitian twister model with the numerically simulated data samples.} Parameters: $\epsilon=0.08$, 
 number of unit cells $N=16$, $m_2=0.6$,   the varying parameter $m_1^{(l)}=0.4106+l*4/\pi^4$ for each sample ${\bf x}^{(l)}$, with $l\in [1,37]$. 
 {\bf a}, Heatmap of the Gaussian kernel values. 
 {\bf b}, Largest eigenvalues of the diffusion matrix $\mathcal{P}$.  
 {\bf c -- f}, Four right eigenvectors $\psi_{0,1,2,3}$ of the diffusion matrix $\mathcal{P}$ with the  corresponding largest four eigenvalues $\lambda_{0,1,2,3} \approx 1$. The horizontal axis denotes the sample number,  and the vertical axis means the values of each sample site in eigenvectors. From the eigenvectors,  we can verify that the jumping points exactly match with the phase transition points $m_1=0.8$ and $m_1=1.6$, indicating  the efficiency of the diffusion map method in classifying non-Hermitian knotted topological phases  with the numerically simulated data samples. 
}
\label{Supp_DM_THE}
\end{figure}

In Fig.~\ref{Supp_DM_THE}, we show the detailed results of learning the non-Hermitian knotted topological phases with the numerically simulated data set.  According to the theory of diffusion map method, the  higher dimensional feature space of each sample can be mapped to the lower dimensional one, with the dimension of the reduced  space is denoted by the number of largest eigenvalues. Then the reduced feature of each sample is mapped to the corresponding component of the right eigenvectors with $\lambda's\approx1$. From  Fig.~\ref{Supp_DM_THE}a, we find that the heatmap is divided into three yellow blocks, indicating the samples can be classified into three categories. In Fig.~\ref{Supp_DM_THE}b, however, there are four eigenvalues approximately equal to 1. This is owing to the fact  that the 29th sample with $m_1\approx1.6015$ is much close to the phase transition point $m_1=1.6$, and causes a large deviation from its presumed category. But it does not influence the general classifying result. One can verify this from the eigenvectors in Fig.~\ref{Supp_DM_THE}(c-f), where the  components  corresponding to the 29th sample are singular. Hence, to plot the three dimensional scatter diagram in Fig.3b of the main text, we omit the special 29th point, and straightforwardly choose the three eigenvectors $\{\psi_0,\psi_2,\psi_3\}$.


\subsection{Experimental result }
\begin{figure}[h]
\centering
\includegraphics[width=0.9\linewidth]{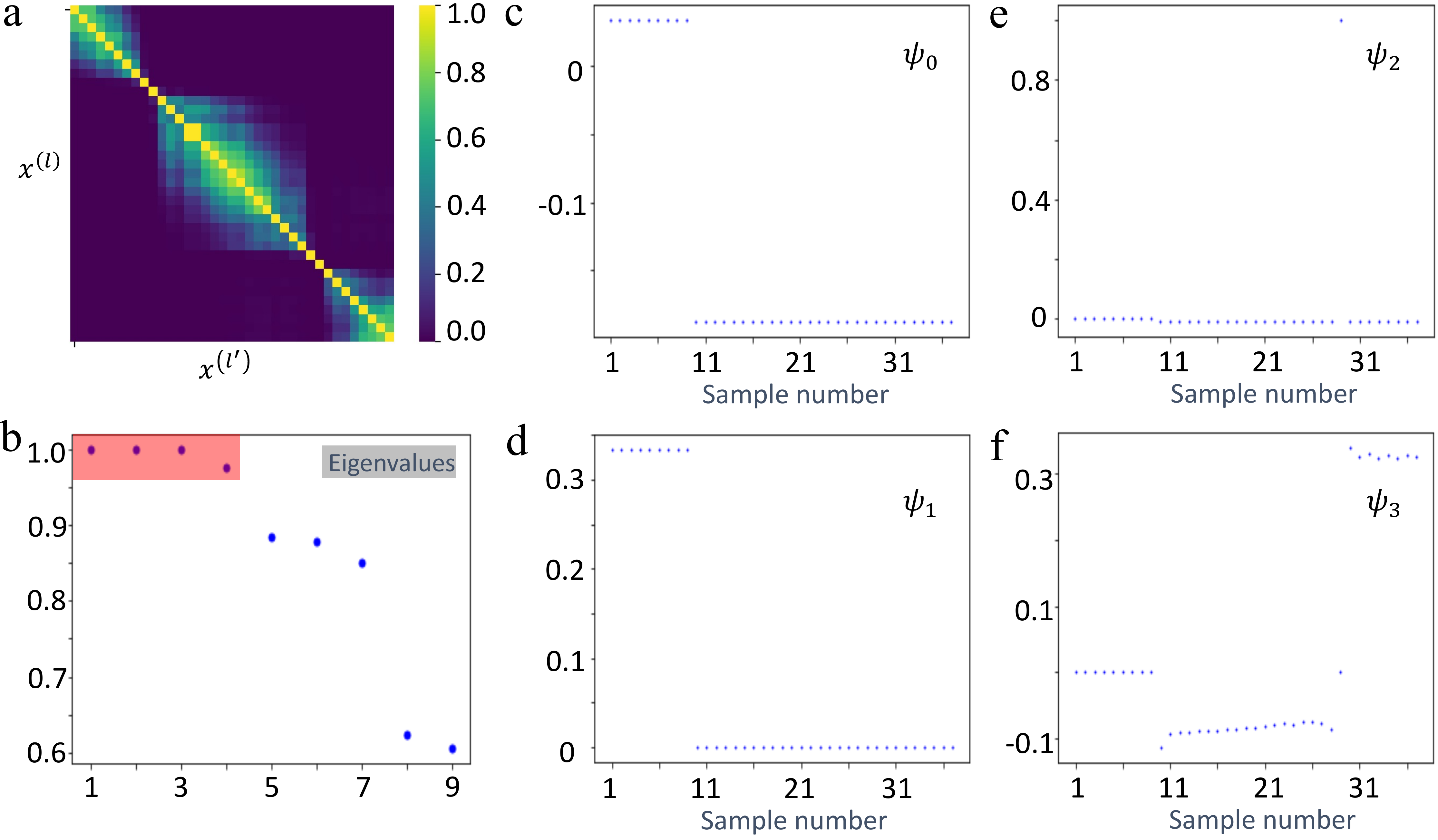}
\caption{{ Classification of 1D Non-Hermitian twister model with the experimental data samples.} Parameter $\epsilon=0.08$. Experimental data samples are implemented by simulating  the Hamiltonian with the following parameters:
 number of unit cells $N=16$, $m_2=0.6$,   the varying parameter $m_1^{(l)}=0.4106+l*4/\pi^4$ for each sample ${\bf x}^{(l)}$, with $l\in [1,37]$. 
 {\bf a}, Heatmap of the Gaussian kernel values. 
 {\bf b}, Largest eigenvalues of the diffusion matrix $\mathcal{P}$.  
 {\bf c -- f}, Four right eigenvectors $\psi_{0,1,2,3}$ of the diffusion matrix $\mathcal{P}$ with the  corresponding largest four eigenvalues $\lambda_{0,1,2,3} \approx 1$. The horizontal axis denotes the sample number,  and the vertical axis means the values of each sample site in eigenvectors. From the eigenvectors,  we can verify that the jumping points exactly match with the phase transition points $m_1=0.8$ and $m_1=1.6$, indicating  the efficiency of the diffusion map method in classifying non-Hermitian knotted topological phases  with the experimental data samples. 
}
\label{Supp_DM_EXP}
\end{figure}

Fig.~\ref{Supp_DM_EXP} demonstrates the detailed results of learning the non-Hermitian knotted topological phases with the experimental data set.  We see from the heatmap in Fig.~\ref{Supp_DM_EXP}a that there are totally three blocks,  indicating that the samples can be classified into three categories. Similar to the numerical simulated case, in Fig.~\ref{Supp_DM_EXP}b,   there are also four eigenvalues approximately equal to 1. As was noted above, this is owing to the  29th sample with $m_1\approx1.6015$,  which is much close to the phase transition point $m_1=1.6$, and causes a large deviation from its presumed category. It does not influence our general classifying result. One can verify this from the eigenvectors in Fig.~\ref{Supp_DM_EXP}(c-f), where the  components  corresponding to the 29th sample are singular. Hence, to plot the three dimensional scatter diagram in Fig.3d of the main text, we omit the special 29th point, and straightforwardly choose the three eigenvectors $\{\psi_0,\psi_1,\psi_3\}$.

\newpage
\begin{table*}[!h]
	\centering
	
	\begin{tabular}{>{\raggedleft}p{30 pt}|>{\raggedleft}p{50 pt}>{\raggedleft}p{50 pt}|>{\raggedleft}p{50 pt}>{\raggedleft\arraybackslash}p{50 pt}|>{\raggedleft}p{50 pt}>{\raggedleft\arraybackslash}p{50 pt}|>{\raggedleft\arraybackslash}p{40 pt}}
		\hline
		\multirow{2}{*}{$k/\pi$}&\multicolumn{2}{c|}{$\langle R_{1}|\sigma_{x}|R_{1}\rangle$}&\multicolumn{2}{c|}{$\langle R_{1}|\sigma_{y}|R_{1}\rangle$}&\multicolumn{2}{c|}{$\langle R_{1}|\sigma_{z}|R_{1}\rangle$}&\multirow{2}{*}{fidelity$/\%$}\\
	    & experiment&theory& experiment&theory& experiment&theory& \\
		\hline
		0.125&0.75(6)&0.796&-0.64(10)&-0.596&0.15(4)&0.102&99.8(9)\\
		0.250&0.56(4)&0.600&-0.82(3)&-0.781&0.09(4)&0.175&99.7(2)\\
		0.375&0.41(3)&0.386&-0.90(1)&-0.898&0.15(3)&0.212&99.9(1)\\
		0.500&0.16(5)&0.168&-0.95(1)&-0.963&0.26(4)&0.209&99.9(2)\\
		0.625&-0.03(24)&-0.048&-0.99(4)&-0.986&0.13(5)&0.157&100.0(24)\\
		0.750&-0.25(5)&-0.258&-0.97(1)&-0.965&-0.04(3)&0.038&99.8(2)\\
		0.875&-0.40(3)&-0.449&-0.88(1)&-0.872&-0.24(3)&-0.194&99.9(1)\\
		1.000&-0.51(3)&-0.596&-0.60(4)&-0.534&-0.61(4)&-0.600&99.7(2)\\
		1.125&-0.65(4)&-0.657&0.05(2)&0.110&-0.75(3)&-0.746&99.9(1)\\
		1.250&-0.48(4)&-0.453&0.52(2)&0.567&-0.71(3)&-0.688&99.9(1)\\
		1.375&-0.10(9)&-0.091&0.73(3)&0.801&-0.68(3)&-0.592&99.7(3)\\
		1.500&0.24(10)&0.315&0.89(4)&0.817&-0.39(6)&-0.484&99.5(7)\\
		1.625&0.78(4)&0.669&0.58(4)&0.646&-0.25(5)&-0.368&99.2(5)\\
		1.750&0.88(2)&0.905&0.31(3)&0.347&-0.35(5)&-0.245&99.7(3)\\
		1.875&0.99(1)&0.993&-0.04(3)&-0.004&-0.12(5)&-0.120&100.0(1)\\
		2.000&0.92(4)&0.943&-0.37(9)&-0.334&0.10(23)&0.000&99.7(23)\\
		\hline
	\end{tabular}
	\caption{\textbf{Experimental data for the eigenstate \bm{$|R_{1}(k)\rangle$} of \bm{$H(k)$} with \bm{$m_{1}=0.5338$} and \bm{$m_{2}=0.6$}}. Theoretically, this chosen parameter falls into the Hopf link phase. The momentum $k$ are equally spaced from 0 to $2\pi$ with an interval of $0.125\pi$. The second, fourth and sixth columns show the experimentally measured data of $\langle \sigma_x\rangle$, $\langle \sigma_y\rangle$ and $\langle \sigma_z\rangle$ respectively. The third, fifth and seventh columns show the theoretical prediction. State fidelities are listed in the last column. The uncertainties are calculated via Monte Carlo simulation for 10,000 times by assuming a Possion distribution of the counts.}
	\label{table1}
\end{table*} 

\begin{table*}[!h]
	\centering
	
	\begin{tabular}{>{\raggedleft}p{30 pt}|>{\raggedleft}p{50 pt}>{\raggedleft}p{50 pt}|>{\raggedleft}p{50 pt}>{\raggedleft\arraybackslash}p{50 pt}|>{\raggedleft}p{50 pt}>{\raggedleft\arraybackslash}p{50 pt}|>{\raggedleft\arraybackslash}p{40 pt}}
		\hline
		\multirow{2}{*}{$k/\pi$}&\multicolumn{2}{c|}{$\langle R_{2}|\sigma_{x}|R_{2}\rangle$}&\multicolumn{2}{c|}{$\langle R_{2}|\sigma_{y}|R_{2}\rangle$}&\multicolumn{2}{c|}{$\langle R_{2}|\sigma_{z}|R_{2}\rangle$}&\multirow{2}{*}{fidelity$/\%$}\\
	    & experiment&theory& experiment&theory& experiment&theory& \\
		\hline
		0.125&-0.99(1)&-0.993&0.05(3)&-0.004&-0.15(5)&-0.120&99.9(1)\\
		0.250&-0.91(2)&-0.905&0.31(3)&0.347&-0.26(6)&-0.245&99.9(2)\\
		0.375&-0.75(3)&-0.669&0.53(3)&0.646&-0.38(5)&-0.368&99.5(3)\\
		0.500&-0.39(4)&-0.315&0.73(2)&0.817&-0.56(3)&-0.484&99.6(2)\\
		0.625&0.04(9)&0.091&0.75(4)&0.801&-0.66(4)&-0.592&99.8(6)\\
		0.750&0.51(9)&0.453&0.53(3)&0.567&-0.67(8)&-0.688&99.9(7)\\
		0.875&0.68(11)&0.657&0.10(2)&0.110&-0.73(12)&-0.746&100.0(9)\\
		1.000&0.53(3)&0.596&-0.53(10)&-0.534&-0.66(8)&-0.600&99.8(6)\\
		1.125&0.49(3)&0.449&-0.85(2)&-0.872&-0.19(3)&-0.194&99.9(1)\\
		1.250&0.29(4)&0.258&-0.96(1)&-0.965&0.01(4)&0.038&100.0(1)\\
		1.375&0.07(14)&0.048&-0.99(2)&-0.986&0.10(4)&0.157&99.9(9)\\
		1.500&-0.20(5)&-0.168&-0.97(1)&-0.963&0.11(4)&0.209&99.7(2)\\
		1.625&-0.43(4)&-0.386&-0.89(2)&-0.898&0.14(4)&0.212&99.8(2)\\
		1.750&-0.51(2)&-0.600&-0.83(2)&-0.781&0.22(4)&0.175&99.7(2)\\
		1.875&-0.77(3)&-0.796&-0.64(4)&-0.596&0.05(3)&0.102&99.9(2)\\
		2.000&-0.92(4)&-0.943&-0.40(19)&-0.334&0.00(30)&0.000&99.9(36)\\
		\hline
	\end{tabular}
	\caption{\textbf{Experimental data for the eigenstate \bm{$|R_{2}(k)\rangle$} of \bm{$H(k)$} with \bm{$m_{1}=0.5338$} and \bm{$m_{2}=0.6$}}. Theoretically, this chosen parameter falls into the Hopf link phase.The experimental data are in good agreement with the theoretical prediction.}
	\label{table2}
\end{table*}

\newpage
\begin{table*}[!h]
	\centering
	
	\begin{tabular}{>{\raggedleft}p{30 pt}|>{\raggedleft}p{50 pt}>{\raggedleft}p{50 pt}|>{\raggedleft}p{50 pt}>{\raggedleft\arraybackslash}p{50 pt}|>{\raggedleft}p{50 pt}>{\raggedleft\arraybackslash}p{50 pt}|>{\raggedleft\arraybackslash}p{40 pt}}
		\hline
		\multirow{2}{*}{$k/\pi$}&\multicolumn{2}{c|}{$\langle R_{1}|\sigma_{x}|R_{1}\rangle$}&\multicolumn{2}{c|}{$\langle R_{1}|\sigma_{y}|R_{1}\rangle$}&\multicolumn{2}{c|}{$\langle R_{1}|\sigma_{z}|R_{1}\rangle$}&\multirow{2}{*}{fidelity$/\%$}\\
	    & experiment&theory& experiment&theory& experiment&theory& \\
		\hline
		0.125&0.44(3)&0.412&-0.85(2)&-0.849&0.30(3)&0.332&100.0(1)\\
		0.250&0.30(3)&0.280&-0.83(2)&-0.828&0.48(3)&0.486&100.0(1)\\
		0.375&0.27(4)&0.169&-0.83(2)&-0.814&0.49(3)&0.556&99.6(2)\\
		0.500&0.11(5)&0.070&-0.81(2)&-0.812&0.57(3)&0.579&100.0(1)\\
		0.625&-0.11(6)&-0.019&-0.84(2)&-0.824&0.53(3)&0.566&99.7(3)\\
		0.750&-0.12(7)&-0.090&-0.85(2)&-0.851&0.51(3)&0.517&100.0(2)\\
		0.875&-0.11(5)&-0.114&-0.92(1)&-0.900&0.37(3)&0.421&99.9(1)\\
		1.000&0.02(7)&0.000&-0.93(1)&-0.943&0.36(2)&0.334&100.0(2)\\
		1.125&0.18(5)&0.114&-0.89(1)&-0.900&0.41(3)&0.421&99.9(2)\\
		1.250&0.12(4)&0.090&-0.86(2)&-0.851&0.50(3)&0.517&100.0(1)\\
		1.375&-0.01(7)&0.019&-0.82(3)&-0.824&0.58(4)&0.566&100.0(2)\\
		1.500&-0.03(6)&-0.070&-0.82(2)&-0.812&0.57(3)&0.579&100.0(2)\\
		1.625&-0.23(3)&-0.169&-0.87(2)&-0.814&0.43(4)&0.556&99.4(3)\\
		1.750&-0.26(3)&-0.280&-0.84(2)&-0.828&0.48(3)&0.486&100.0(1)\\
		1.875&-0.41(3)&-0.412&-0.84(2)&-0.849&0.36(4)&0.332&100.0(1)\\
		2.000&-0.52(3)&-0.606&-0.85(2)&-0.796&0.09(10)&0.000&99.5(6)\\
		\hline
	\end{tabular}
	\caption{\textbf{Experimental data for the eigenstate \bm{$|R_{1}(k)\rangle$} of \bm{$H(k)$} with \bm{$m_{1}=1.2730$} and \bm{$m_{2}=0.6$}}. Theoretically, this chosen parameter falls into the unknot phase. The experimental data are in good agreement with the theoretical prediction.}
	\label{table3}
\end{table*}

\begin{table*}[!h]
	\centering
	
	\begin{tabular}{>{\raggedleft}p{30 pt}|>{\raggedleft}p{50 pt}>{\raggedleft}p{50 pt}|>{\raggedleft}p{50 pt}>{\raggedleft\arraybackslash}p{50 pt}|>{\raggedleft}p{50 pt}>{\raggedleft\arraybackslash}p{50 pt}|>{\raggedleft\arraybackslash}p{40 pt}}
		\hline
		\multirow{2}{*}{$k/\pi$}&\multicolumn{2}{c|}{$\langle R_{2}|\sigma_{x}|R_{2}\rangle$}&\multicolumn{2}{c|}{$\langle R_{2}|\sigma_{y}|R_{2}\rangle$}&\multicolumn{2}{c|}{$\langle R_{2}|\sigma_{z}|R_{2}\rangle$}&\multirow{2}{*}{fidelity$/\%$}\\
	    & experiment&theory& experiment&theory& experiment&theory& \\
		\hline
		0.125&-0.82(2)&-0.832&-0.44(3)&-0.433&-0.37(4)&-0.348&100.0(1)\\
		0.250&-0.84(2)&-0.841&0.00(3)&0.031&-0.54(3)&-0.541&100.0(1)\\
		0.375&-0.61(5)&-0.632&0.40(3)&0.402&-0.69(4)&-0.663&100.0(2)\\
		0.500&-0.39(5)&-0.290&0.55(3)&0.592&-0.74(3)&-0.752&99.7(3)\\
		0.625&0.20(7)&0.079&0.61(3)&0.563&-0.76(2)&-0.823&99.5(5)\\
		0.750&0.44(7)&0.350&0.32(3)&0.320&-0.84(4)&-0.880&99.7(4)\\
		0.875&0.40(7)&0.378&-0.10(3)&-0.070&-0.91(3)&-0.923&100.0(2)\\
		1.000&0.02(12)&0.000&-0.38(4)&-0.343&-0.92(2)&-0.939&100.0(7)\\
		1.125&-0.42(5)&-0.378&-0.08(3)&-0.070&-0.90(2)&-0.923&99.9(2)\\
		1.250&-0.42(5)&-0.350&0.39(3)&0.320&-0.82(3)&-0.880&99.7(3)\\
		1.375&-0.11(8)&-0.079&0.54(3)&0.563&-0.84(2)&-0.823&99.9(3)\\
		1.500&0.28(8)&0.290&0.64(4)&0.592&-0.72(4)&-0.752&99.9(3)\\
		1.625&0.64(8)&0.632&0.38(4)&0.402&-0.67(7)&-0.663&100.0(3)\\
		1.750&0.83(2)&0.841&0.01(3)&0.031&-0.55(3)&-0.541&100.0(1)\\
		1.875&0.83(2)&0.832&-0.43(3)&-0.433&-0.35(3)&-0.348&100.0(1)\\
		2.000&0.56(3)&0.606&-0.83(2)&-0.796&-0.02(10)&0.000&99.9(5)\\
		\hline
	\end{tabular}
	\caption{\textbf{Experimental data for the eigenstate \bm{$|R_{2}(k)\rangle$} of \bm{$H(k)$} with \bm{$m_{1}=1.2730$} and \bm{$m_{2}=0.6$}}. Theoretically, this chosen parameter falls into the unknot phase. The experimental data are in good agreement with the theoretical prediction.}
	\label{table4}
\end{table*}

\newpage
\begin{table*}[!h]
	\centering
	
	\begin{tabular}{>{\raggedleft}p{30 pt}|>{\raggedleft}p{50 pt}>{\raggedleft}p{50 pt}|>{\raggedleft}p{50 pt}>{\raggedleft\arraybackslash}p{50 pt}|>{\raggedleft}p{50 pt}>{\raggedleft\arraybackslash}p{50 pt}|>{\raggedleft\arraybackslash}p{40 pt}}
		\hline
		\multirow{2}{*}{$k/\pi$}&\multicolumn{2}{c|}{$\langle R_{1}|\sigma_{x}|R_{1}\rangle$}&\multicolumn{2}{c|}{$\langle R_{1}|\sigma_{y}|R_{1}\rangle$}&\multicolumn{2}{c|}{$\langle R_{1}|\sigma_{z}|R_{1}\rangle$}&\multirow{2}{*}{fidelity$/\%$}\\
	    & experiment&theory& experiment&theory& experiment&theory& \\
		\hline
		0.125&0.15(4)&0.131&-0.85(2)&-0.775&0.51(3)&0.619&99.6(3)\\
		0.250&0.14(3)&0.121&-0.77(3)&-0.711&0.63(3)&0.693&99.8(2)\\
		0.375&0.13(3)&0.080&-0.66(2)&-0.679&0.74(2)&0.730&99.9(1)\\
		0.500&0.02(7)&0.034&-0.65(4)&-0.667&0.76(3)&0.744&100.0(2)\\
		0.625&0.14(11)&-0.009&-0.67(5)&-0.672&0.73(4)&0.740&99.4(13)\\
		0.750&0.05(7)&-0.039&-0.72(4)&-0.693&0.69(4)&0.720&99.7(5)\\
		0.875&0.03(9)&-0.040&-0.77(3)&-0.724&0.63(3)&0.689&99.7(6)\\
		1.000&0.04(8)&0.000&-0.79(3)&-0.742&0.61(3)&0.670&99.8(5)\\
		1.125&0.07(6)&0.40&-0.77(2)&-0.724&0.63(3)&0.689&99.8(2)\\
		1.250&0.00(5)&0.039&-0.69(3)&-0.693&0.72(3)&0.720&100.0(2)\\
		1.375&-0.16(8)&0.009&-0.68(4)&-0.672&0.72(3)&0.740&99.3(8)\\
		1.500&-0.04(6)&-0.034&-0.64(4)&-0.667&0.77(3)&0.744&100.0(2)\\
		1.625&-0.06(4)&-0.080&-0.69(4)&-0.679&0.73(3)&0.730&100.0(1)\\
		1.750&-0.19(4)&-0.121&-0.76(3)&-0.711&0.63(4)&0.693&99.7(2)\\
		1.875&-0.14(4)&-0.131&-0.81(2)&-0.775&0.57(3)&0.619&99.9(1)\\
		2.000&-0.01(13)&0.000&-0.85(2)&-0.847&0.53(3)&0.532&100.0(7)\\
		\hline
	\end{tabular}
	\caption{\textbf{Experimental data for the eigenstate \bm{$|R_{1}(k)\rangle$} of \bm{$H(k)$} with \bm{$m_{1}=1.8889$} and \bm{$m_{2}=0.6$}}. Theoretically, this chosen parameter falls into the unlink phase. The experimental data are in good agreement with the theoretical prediction.}
	\label{table5}
\end{table*}

\begin{table*}[!h]
	\centering
	
	\begin{tabular}{>{\raggedleft}p{30 pt}|>{\raggedleft}p{50 pt}>{\raggedleft}p{50 pt}|>{\raggedleft}p{50 pt}>{\raggedleft\arraybackslash}p{50 pt}|>{\raggedleft}p{50 pt}>{\raggedleft\arraybackslash}p{50 pt}|>{\raggedleft\arraybackslash}p{40 pt}}
		\hline
		\multirow{2}{*}{$k/\pi$}&\multicolumn{2}{c|}{$\langle R_{2}|\sigma_{x}|R_{2}\rangle$}&\multicolumn{2}{c|}{$\langle R_{2}|\sigma_{y}|R_{2}\rangle$}&\multicolumn{2}{c|}{$\langle R_{2}|\sigma_{z}|R_{2}\rangle$}&\multirow{2}{*}{fidelity$/\%$}\\
	    & experiment&theory& experiment&theory& experiment&theory& \\
		\hline
		0.125&-0.60(4)&-0.561&-0.64(4)&-0.538&-0.48(2)&-0.630&99.1(4)\\
		0.250&-0.75(3)&-0.679&-0.06(3)&-0.083&-0.66(3)&-0.729&99.7(2)\\
		0.375&-0.54(3)&-0.532&0.31(3)&0.276&-0.78(2)&-0.800&99.9(1)\\
		0.500&-0.29(11)&-0.244&0.42(4)&0.457&-0.86(4)&-0.855&99.9(6)\\
		0.625&0.10(6)&0.064&0.38(3)&0.431&-0.92(2)&-0.900&99.9(2)\\
		0.750&0.27(13)&0.269&0.23(4)&0.223&-0.94(4)&-0.937&100.0(10)\\
		0.875&0.32(12)&0.256&-0.14(4)&-0.066&-0.94(4)&-0.965&99.8(8)\\
		1.000&-0.07(13)&0.000&-0.19(4)&-0.220&-0.98(2)&-0.976&99.8(10)\\
		1.125&-0.28(9)&-0.256&-0.07(3)&-0.066&-0.96(3)&-0.965&100.0(3)\\
		1.250&-0.36(7)&-0.269&0.18(3)&0.223&-0.92(3)&-0.937&99.7(4)\\
		1.375&-0.09(7)&-0.064&0.47(4)&0.431&-0.88(2)&-0.900&99.9(2)\\
		1.500&0.21(15)&0.244&0.46(5)&0.457&-0.86(4)&-0.855&100.0(17)\\
		1.625&0.54(8)&0.532&0.22(4)&0.276&-0.81(5)&-0.800&99.9(3)\\
		1.750&0.76(3)&0.679&-0.04(3)&-0.083&-0.65(3)&-0.729&99.7(2)\\
		1.875&0.53(5)&0.561&-0.57(4)&-0.538&-0.62(3)&-0.630&99.9(2)\\
		2.000&0.08(7)&0.000&-0.87(2)&-0.847&-0.49(3)&-0.532&99.8(3)\\
		\hline
	\end{tabular}
	\caption{\textbf{Experimental data for the eigenstate \bm{$|R_{2}(k)\rangle$} of \bm{$H(k)$} with \bm{$m_{1}=1.8889$} and \bm{$m_{2}=0.6$}}. Theoretically, this chosen parameter falls into the unlink phase. The experimental data are in good agreement with the theoretical prediction.}
	\label{table6}
\end{table*}




\end{document}